# Deep dip teardown of tubeless insulin pump

Sergei Skorobogatov
Computer Laboratory
University of Cambridge
Cambridge, UK
sps32@cam.ac.uk

***Abstract*—This paper introduces a deep level teardown process of a personal medical device – the OmniPod® wireless tubeless insulin pump. This starts with mechanical teardown exposing the engineering solutions used inside the device. Then the electronic part of the device is analysed followed by components identification. Finally, the firmware extraction is performed allowing further analysis of the firmware inside the device as well as real-time debugging. This paper also evaluates the security of the main controller IC of the device. It reveals some weaknesses in the device design process which lead to the possibility of the successful teardown. Should the hardware security of the controller inside the device was well thought through, the teardown process would be far more complicated. This paper demonstrates what the typical teardown process of a personal medical device involves. This knowledge could help in improving the hardware security of sensitive devices.**

## I. Introduction

Security of medical devices is a very important issue because any failures could have a drastic effect on the patients' health [1]. Many modern wearable medical devices have complex semiconductor chips inside – either microcontroller or SoC (System-on-Chip). Those semiconductor chips are supposed to maintain both integrity and confidentiality of the information stored inside. This should prevent firmware analysis and modification in order to hijack control over the device. As many medical devices keep logging patient's data, their security is paramount. Eavesdropping on communication channel and taking over the device control must also be prevented to maintain confidentiality of data and avoid any physical harm to patients.

In the past there were several publications and reports on the security flaws in personal medical devices. This involves vulnerabilities of implantable cardiac defibrillators [2] and insulin pumps [3]. The most important aspect of any insulin pump is its security. This is because the insulin delivery process must be carefully adjusted to the patient's blood glucose level (also called blood sugar level). Should the glucose level become too low this causes an unpleasant symptoms called hypoglycaemia. If the glucose level becomes too high the patient experience hyperglycaemia. Both conditions are quite dangerous and could result in a coma or even death of a patient. Respectively, delaying the injection of insulin after taking a food is likely to result in hyperglycaemia, while injecting too much of insulin is likely to cause hypoglycaemia. Because modern insulin pumps are controlled wirelessly, hijacking of the communication would give a potential attacker possibility to interfere with the delivery of insulin. This could cause a serious harm to the patient's health.

This paper analyses a particular wireless tubeless insulin pump – the OmniPod® device manufactured by Insulet [4]. This is the only personal tubeless insulin pump approved by FDA (US Food and Drug Administration) [5].

This paper is organised as follows. Section 2 gives brief introduction to insulin pumps. The mechanical teardown process is presented in Section 3, followed by the description of the electrical circuit teardown in Section 4. Components identification and analysis are described in Section 5. Firmware extraction process is outlined in Section 6. The approach to further code analysis is described in Section 7. This is followed by discussions and future work outlined in Section 8 and the conclusion in Section 9.

## II. Background

People with Type 1 diabetes are relying on constant delivery of insulin into the blood of their body [6]. In the old days this was achieved with multiple syringe injections or jabs of insulin throughout a day. With the development of insulin pumps this process was automated with portable devices attached to the body. The pump can continuously deliver amounts of rapid or short acting insulin via a catheter placed under the skin [7]. This reduces the need for multiple insulin jabs per day and gives the user increased ability to control blood glucose levels. An insulin pump consists of the main pump unit with an insulin reservoir attached to a long, thin piece of tubing with a needle or cannula at one end. These tubes pose a significant drawback for the wide use of insulin pumps – they have tubes hanging around patient's body. Not only these tubes can be trapped by clothes or furniture, but the pumps are not waterproof. A significant improvement came with the development of tubeless insulin pumps [8]. However, they still have some drawbacks. In particular, they do not have direct integration with a Continuous Glucose Monitor (CGM).

An informal community of developers has taken it upon themselves to develop open source insulin delivery systems that would improve the lives of patients with Type 1 diabetes. Examples of these systems include Loop [9] and OpenAPS [10]. The work is also supported by the Nightscout Foundation [11]. There was a lot of media coverage on the artificial pancreas project in the past [12]. Such projects pose a lot of challenges because the manufacturers of insulin pumps do not provide much information about their devices. Hence, the work of such communities and groups involves a lot of hacking and reverse engineering. Insulin pump electronics usually contain a single chip that performs authentication and control. So far they were only able to understand and use the conventional pump made by Medtronic and only for certain older versions of the firmware. For a long time they tried to understand the communication protocol of OmniPod devices, but without much success [13]. However, they do have a dedicated page where all their findings are shared [14].

The OmniPod system consists of the Pod that holds the insulin dose and the PDM (Personal Diabetes Manager) that communicates wirelessly with the Pod to deliver continuous insulin based on patient's personal settings (Figure 1).

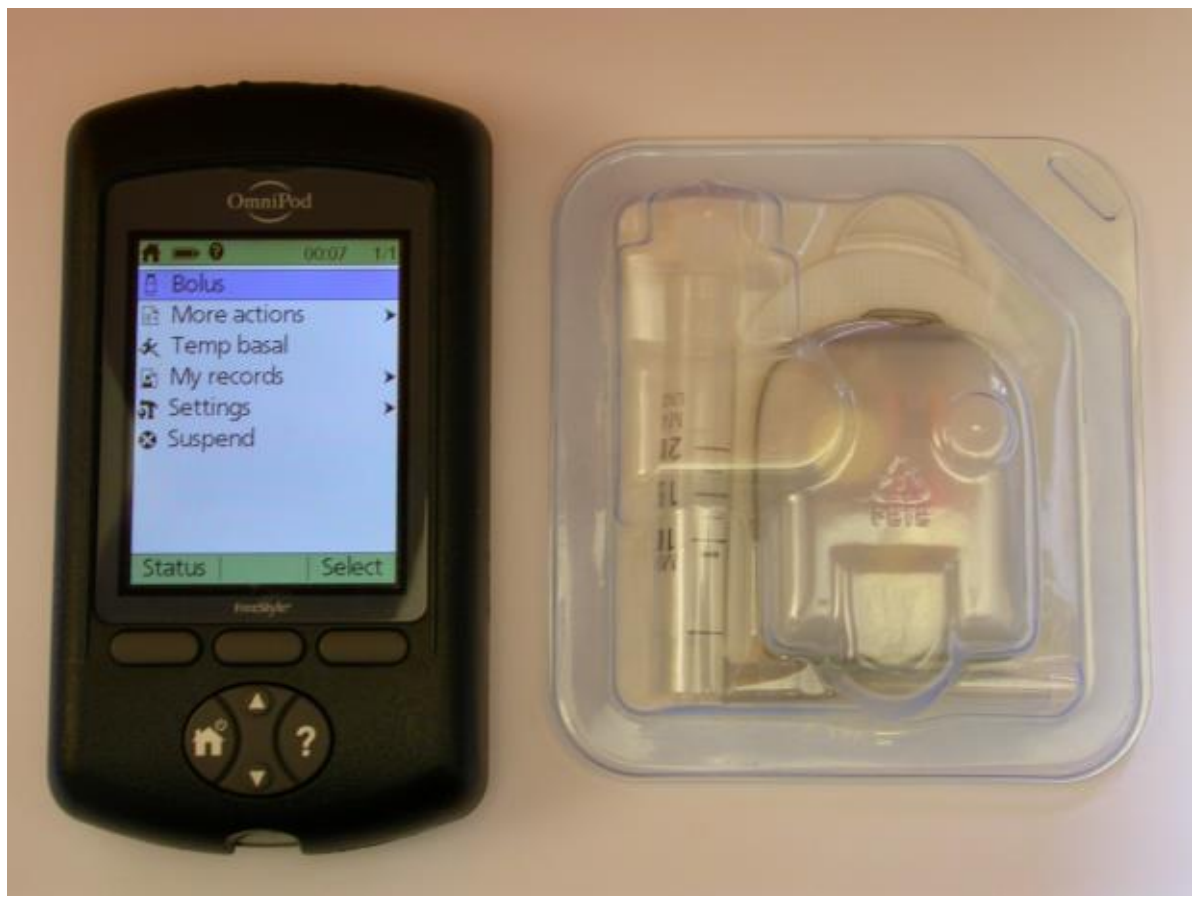

Fig. 1. Insulet OmniPod system: PDM and Pod.

Figure 2 shows what the Pod looks like after taking it out of the sterilised packaging.

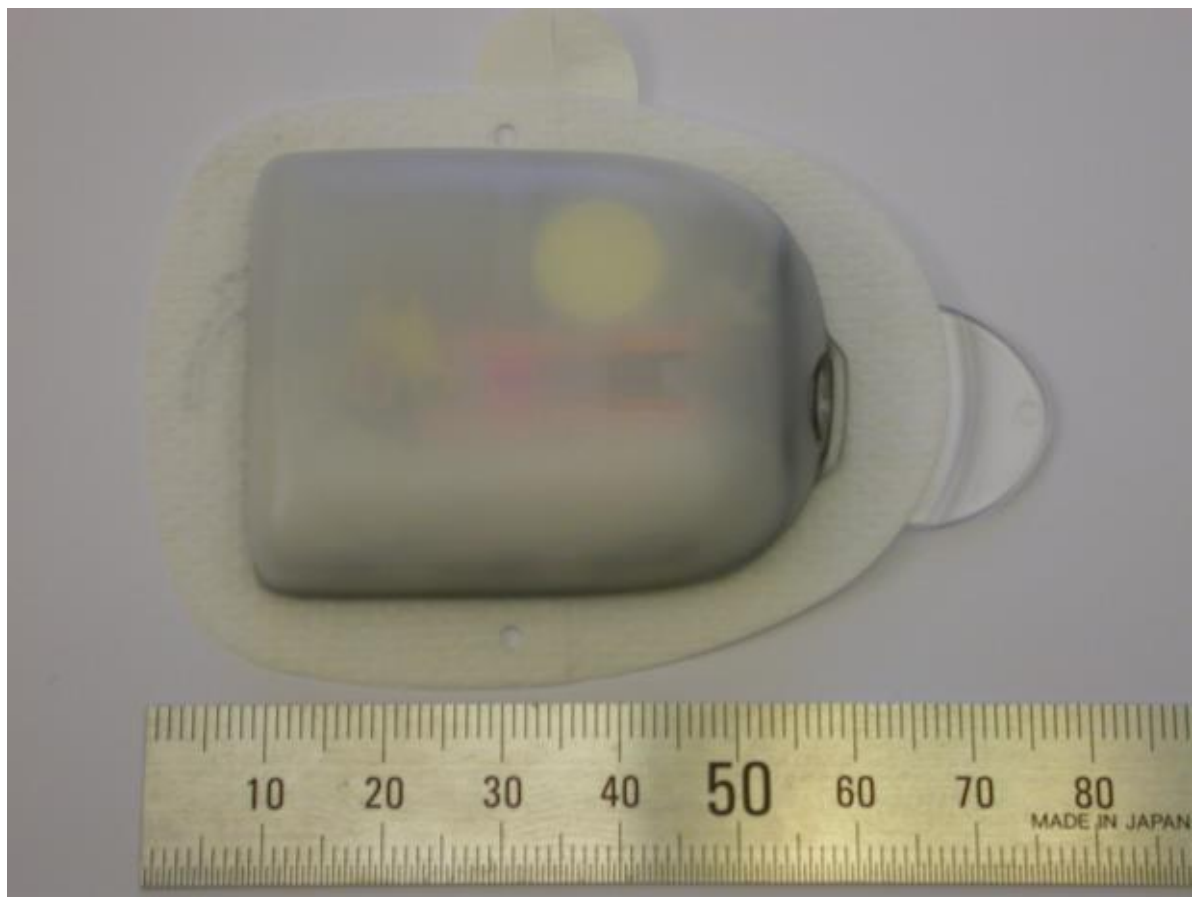

Fig. 2. Unused Pod.

The Pod must be initialised first before the use. This involves filling it with insulin, attaching to the body and activating the insertion of a cannula under the skin.

## III. Mechanical teardown

The purpose of the mechanical teardown is to understand how the device works and what engineering solutions were used for building it. Very often medical devices have multiple patents on the solutions used in their design and manufacturing. This could help in better understanding their functionality and purposes of some parts.

### A. *Opening up the Pod*

The Pod's top cover is so strongly glued to the base that it was necessary to use an engraving tool with a small circular saw attached to it. The steel saw bit was about 20mm in diameter and 0.3mm thin. It easily cuts through the plastic and it is the black plastic in between the opaque top and transparent base that is needed to be cut through. After the cutting is finished along the perimeter, the top and bottom parts can be easily detached from each other (Figure 3).

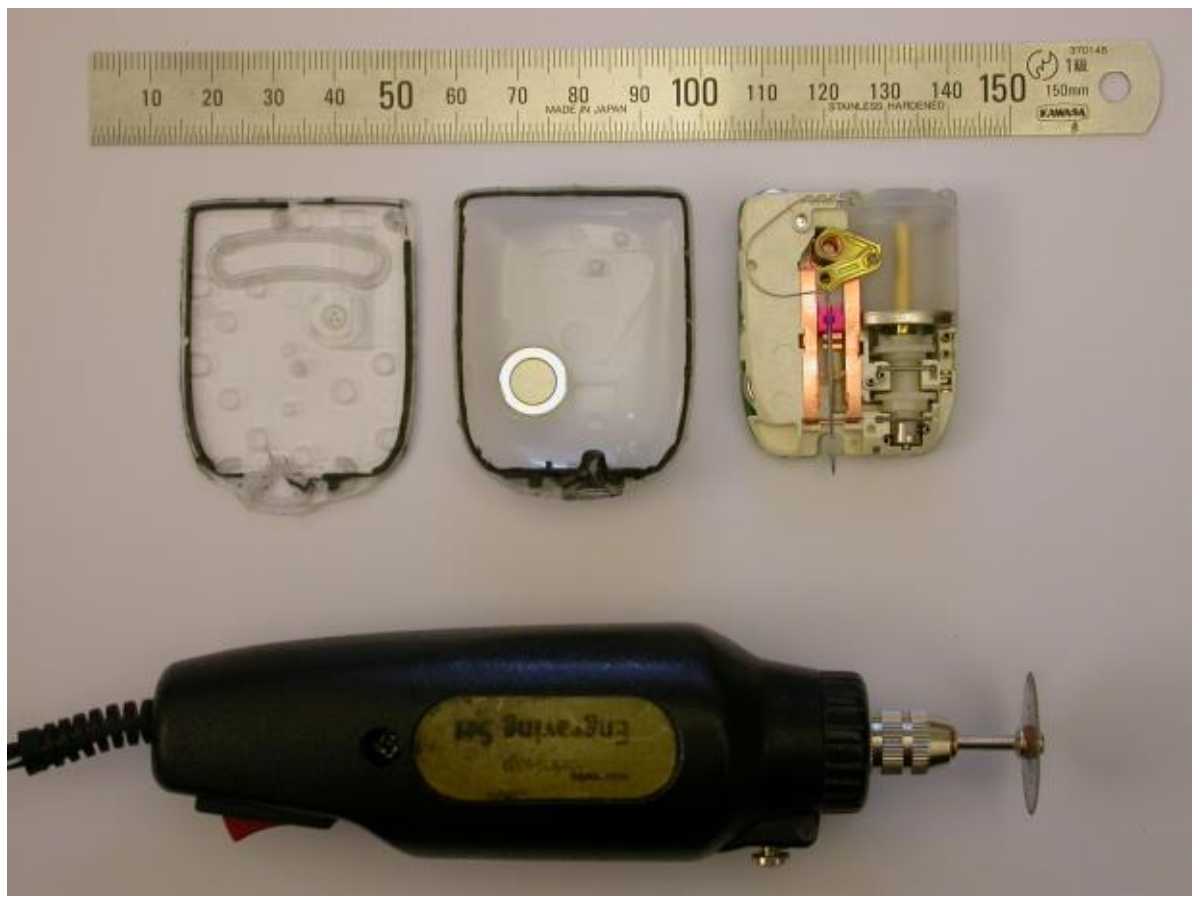

Fig. 3. Pod with its cover and base detached.

The top cover holds a piezo buzzer that produces some beeps during the Pod operation to attract patient's attention. The pictures of the internal assembly from both sides are presented in Figures 4 and 5.

Inside the assembly there is a reservoir for insulin supply of up to 3 days of life of an activated Pod. In the middle there is a spring-loaded cannula insertion mechanism. The white motor gears can be seen next to the reservoir. They are driven by a motor mechanism at the back side of the assembly.

The PCB (Printed Circuit Board) with all the electronic components is attached to the back side of the assembly. It is held in place by six plastic bumps which can be easily removed with a sharp knife. There are three 1.5V batteries which supply the Pod's electronics. They are held in place by metal springs for better electrical contact. Other small springs are used to connect the PCB with other parts of the assembly. Two springs in the middle establish a contact with piezo buzzer. At the end of the white plastic gears there is an encoder connected to the PCB with two springs. Other two springs next to the reservoir are used to monitor the position of the membrane.

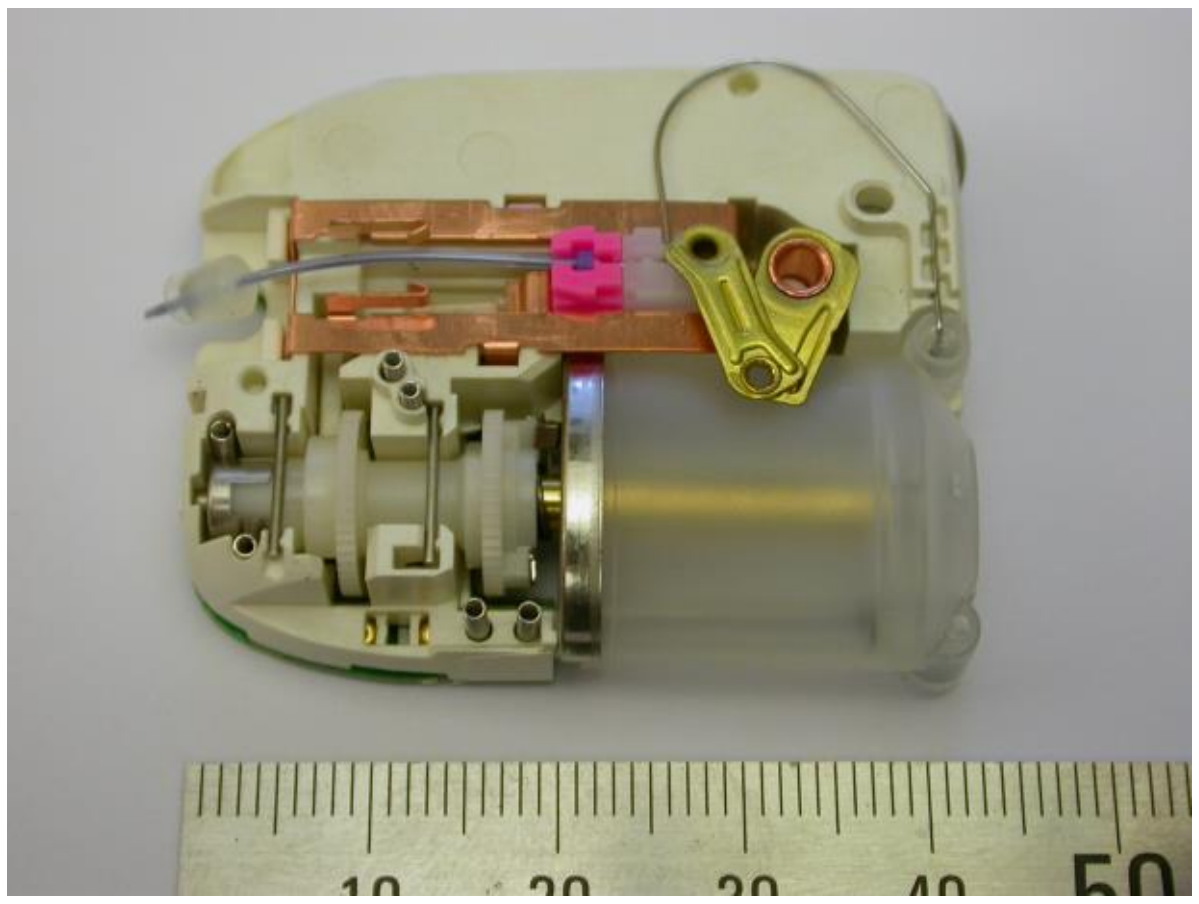

Fig. 4. Top view on the pump assembly inside the Pod.

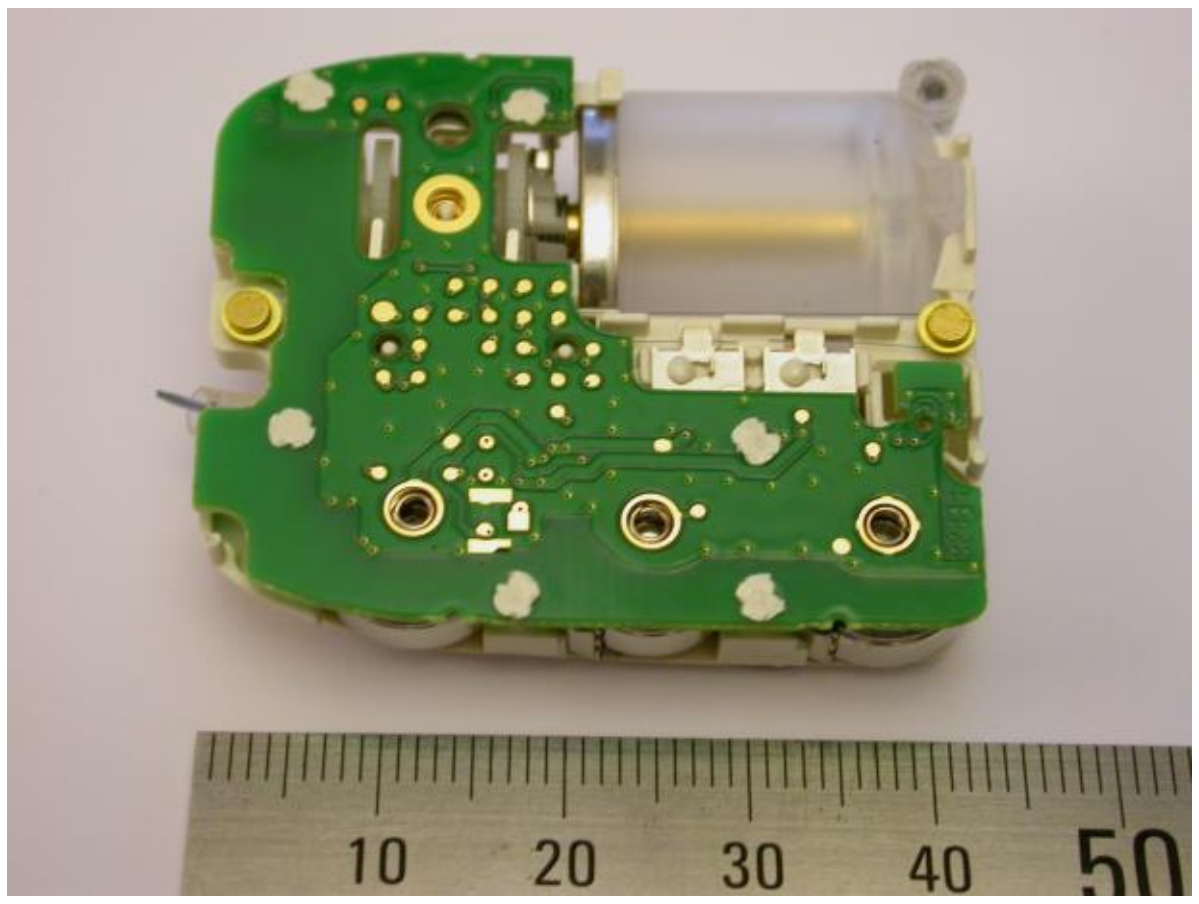

Fig. 5. Bottom view on the pump assembly inside the Pod.

## *B. How the stuff works*

Before any use the new Pod must be activated. This is achieved with placing it next to the PDM and following the activation instructions on the screen. When ready the PDM asks for the Pod to be filled with insulin. During the reservoir filling process the membrane inside it pushes the steel rod forward. Once the rod shorts the two springs outside the reservoir the Pod makes two beeps and also tells PDM that it is ready for use. However, as this happens at about 1/3 of the capacity, it can be filled further as much as needed for up to three days of insulin supply.

The next step is pairing of the Pod to the PDM (Figure 6). During this step the PDM establishes a way for secure communication with the Pod. It also activates the motor inside the Pod and this can be observed with the move of the motor gears. At one side of the gears there is a locking spring which is activated with the turn of the gear. Before the activation happen the brass rod was not held by the spring, hence, the membrane was free to move allowing the reservoir to be filled with insulin. Once released, the spring acts as a clutch connecting the gears to the threaded brass rod inside the reservoir. From this moment the gears will turn the rod and push the membrane forward. That way the insulin will be pushed out of the reservoir into the metal tube and then through the plastic cannula inside the patient’s body.

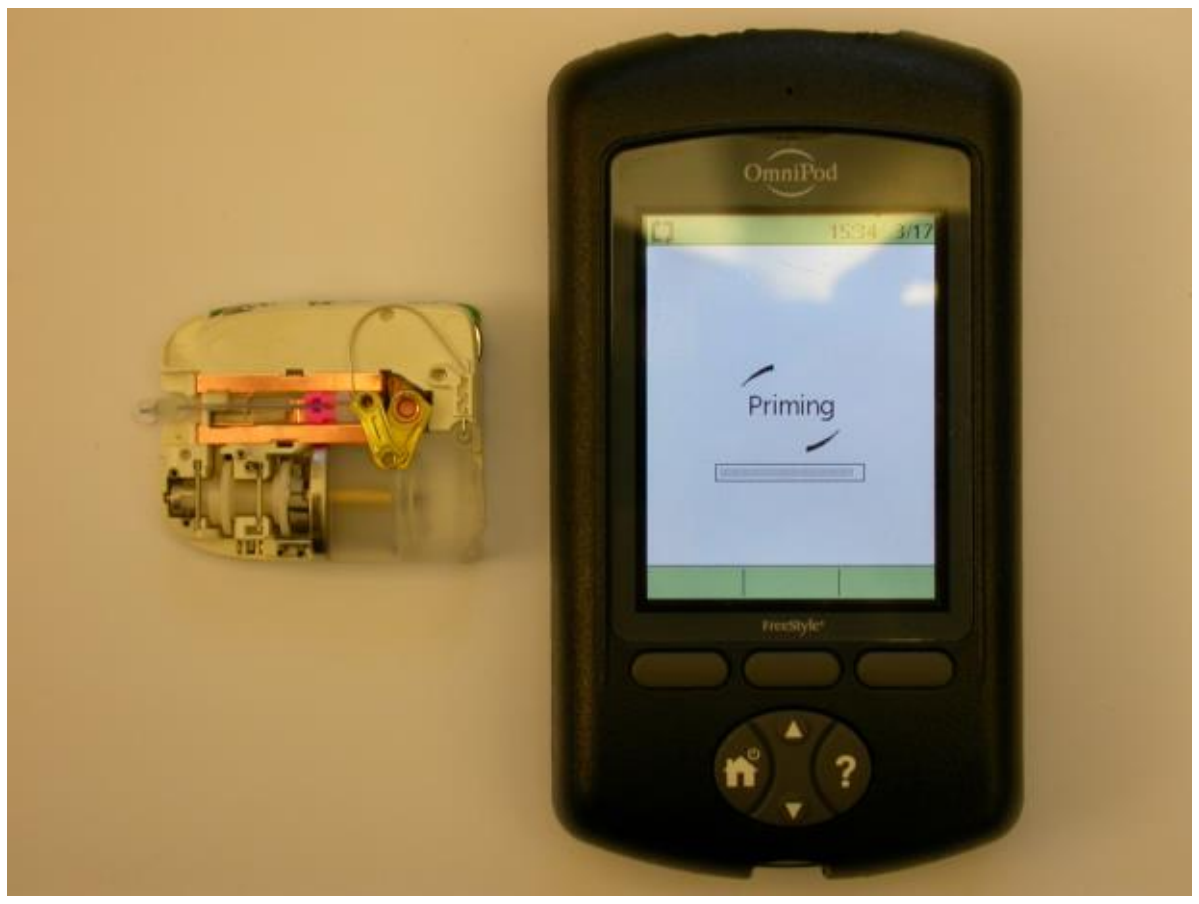

Fig. 6. Pairing the Pod to PDM.

After the priming is finished the PDM is ready to activate the Pod and insert the cannula. A few more turns of the gears release the plastic lever that holds the strong spring loaded insertion mechanism. The spring pushes the metal needle together with the plastic cannula fast forward with a force and then quickly retracts the needle leaving the cannula inside the skin. The result of this operation on the opened Pod is shown in Figure 7.

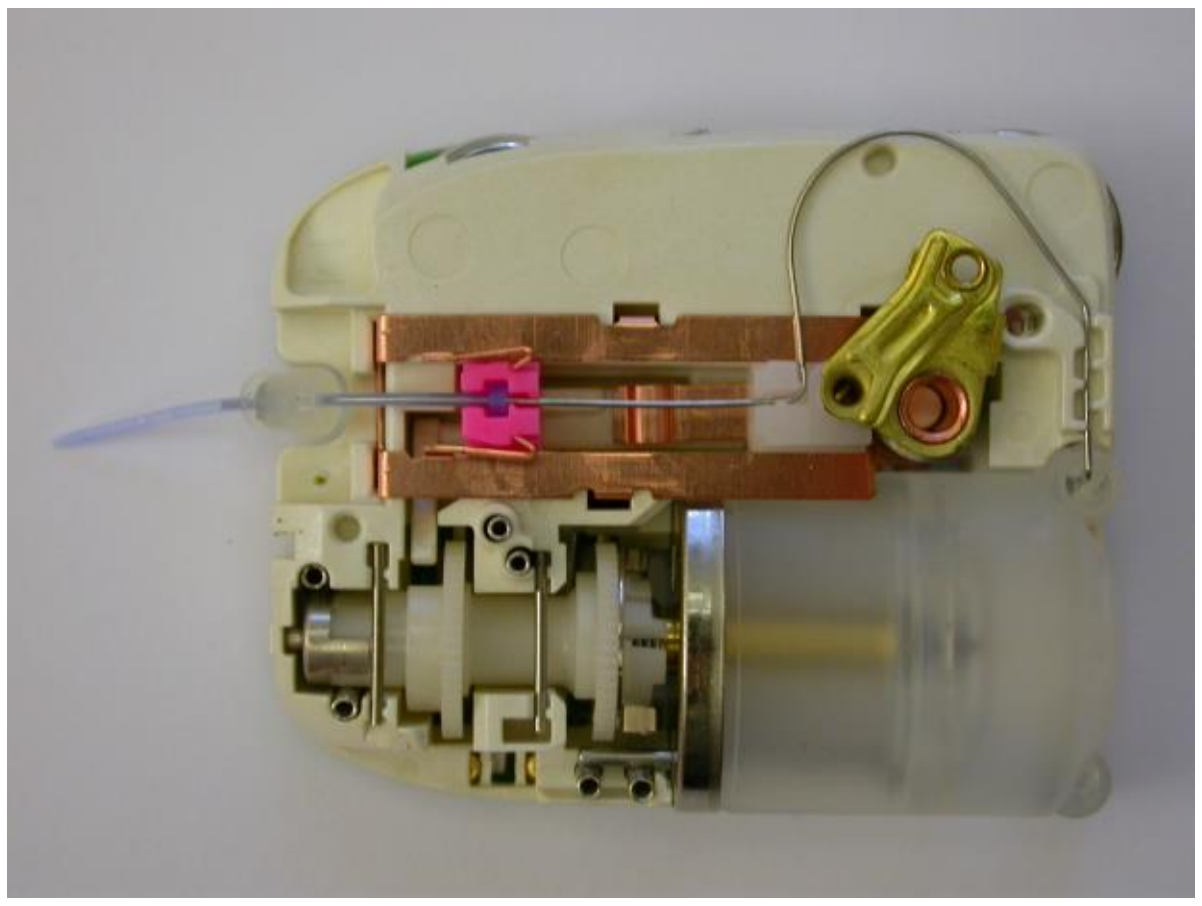

Fig. 7. Opened Pod after cannula insertion.

The motor mechanism of the Pod can be seen when the PCB is detached from the assembly (Figure 8). The gears are driven by an anchor mechanism pushed with a muscle wires. These wires become shorter when an electric current is sent through them and they are heated [15]. The anchor is connected to the ground potential, while either of the two terminals are connected to the power supply to send the current through the corresponding wire. On the far end of the anchor there is a pin which connects to one of the springs once the anchor reaches the desired position. That way the controller knows when to switch off the current to prevent overheating and excessive battery draw.

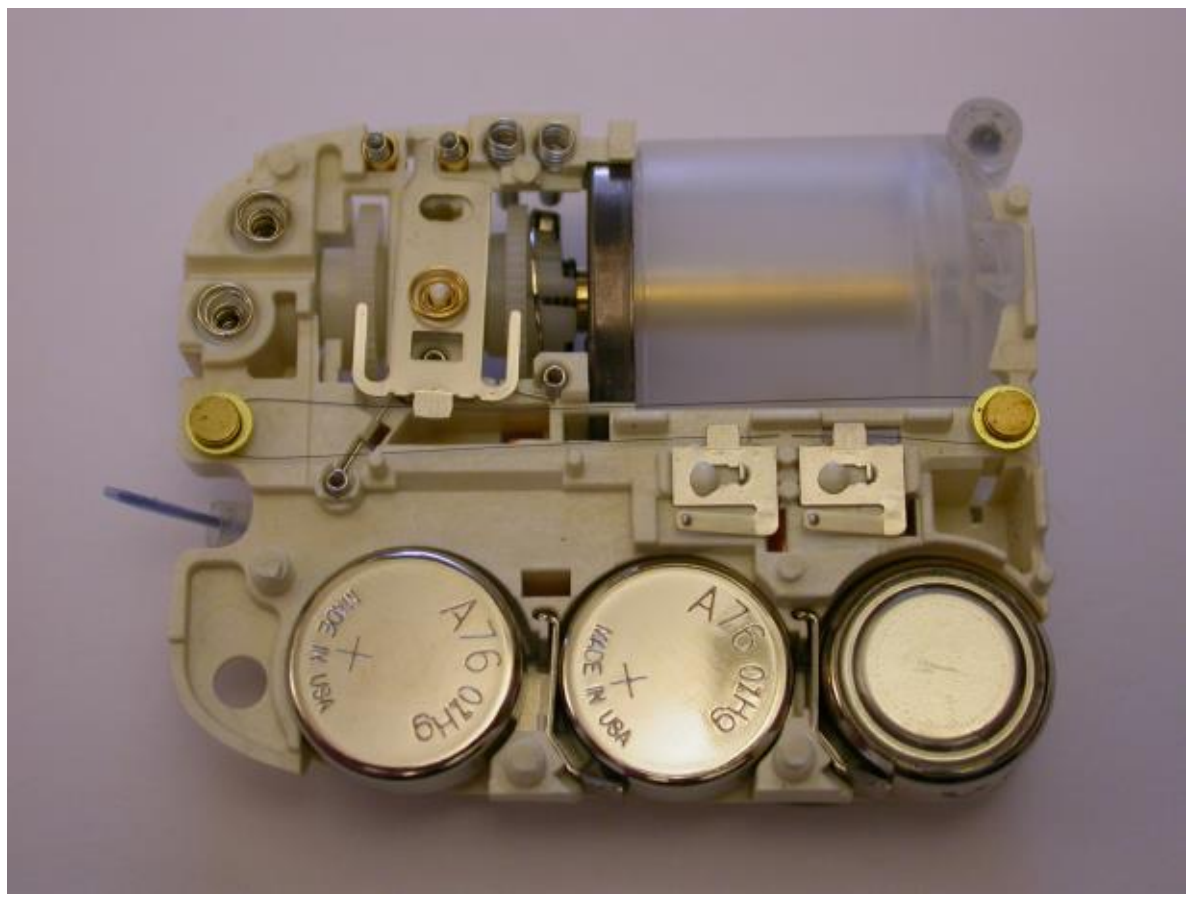

Fig. 8. Bottom side of the Pod after PCB is removed.

The video of the whole process of opening the Pod and demonstration of its operation is available online [16].

In order to better understand the functionality of the device we need to know both its electrical schematic and the software it runs.

### C. Opening up the PDM

The PDM that controls the Pod was also opened to investigate its internal components (Figure 9).

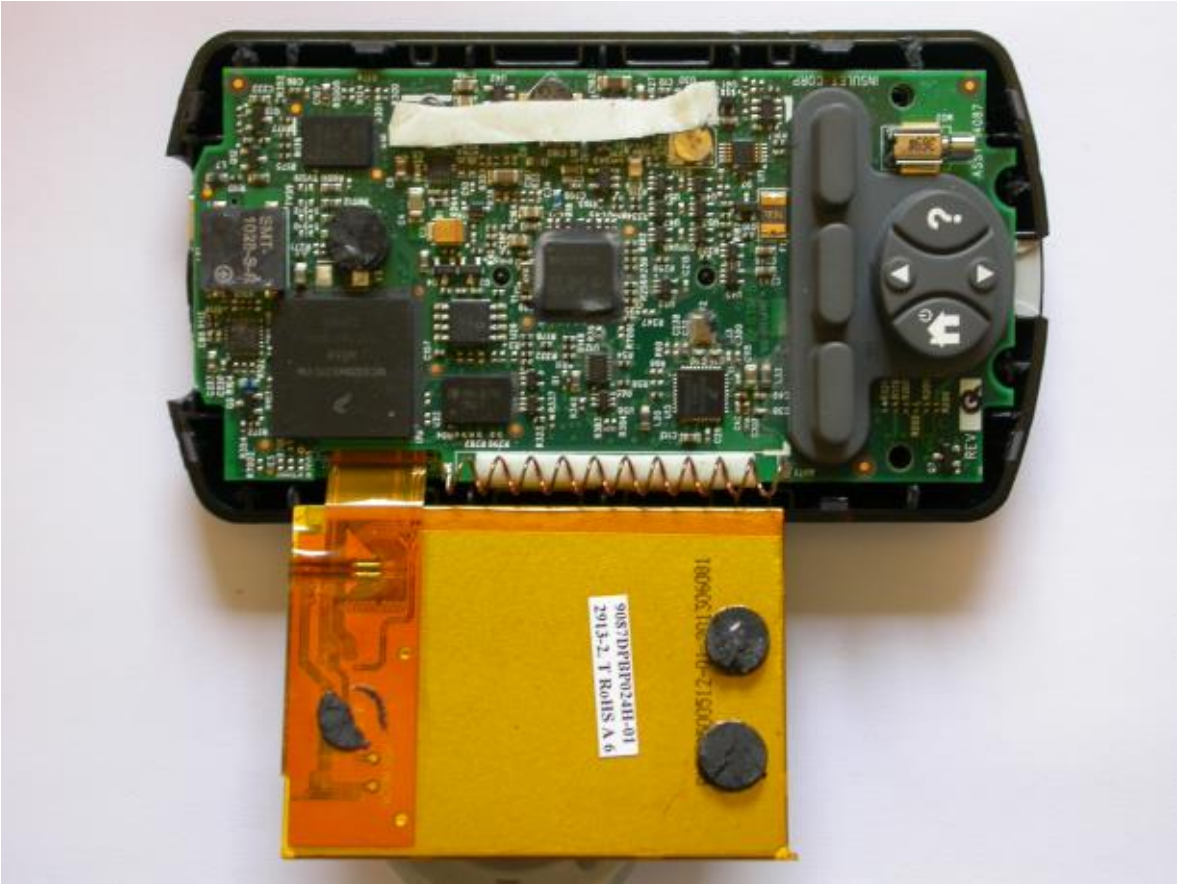

Fig. 9. Opened up PDM.

There is a large ARM-based microprocessor inside it (Freescale MC9328MX21SVM) that runs the control software. The communication is performed by the same type of microcontroller found in the Pod – Freescale SoC S9S8ER48 or its newer version SC9S08ER48. Freescale merged with NXP in 2015, therefore, any official information on the components can only be found from NXP website.

Some analysis work has already been carried out by OpenAPS community on the main controller inside PDM. However, it turned out that all the handling of communication packets was performed by the S9S8ER48/SC9S08ER48 SoC controller. Since all Pods are discarded after use, there were no issues in obtaining as many samples of those SoC chips as needed. Therefore, all the efforts were moved towards reverse engineering of the Pod.

Table 1 outlines different versions of the controller chip found in various Pods and PDMs .

TABLE I. VERSIONS OF 9S08ER48 CHIP FOUND IN DEVICES

| Device | Chip Name | Silicon Revision | Firmware Revision | Chip Date |
|---|---|---|---|---|
| Pod | S9S8ER48 | 1N80A | 14549J | 09/2012 |
| Pod | SC9S08ER48CHP | 2N80A | 16148A | 11/2013 |
| Pod | SC9S08ER48CHP | 2N80A | 16148C | 09/2014 |
| Pod | SC9S08ER48CHP | 2N80A | 16533B | 02/2015 |
| PDM | S9S8ER48 | 1N80A | Unknown | 03/2013 |
| PDM | SC9S08ER48CHP | 1N80A | Unknown | 11/2013 |
| PDM | SC9S08ER48CHP | 2N80A | 16544A | 08/2016 |

For this research only the latest firmware revision in the Pod was analysed marked on the chip as 16533B.

## IV. CIRCUIT LEVEL TEARDOWN

All the electronic components of the Pod are placed on the PCB attached to the assembly. The picture of the detached PCB is presented in Figure 10. The IC in 40-pin QFN package is the only integrated circuit on the board. The IC has the Freescale logo and the rest of the marking is: SC9S08ER, 48CHP, 2N80A, CTLJ603A. Unfortunately, it was not possible to find any information about this chip on the NXP website (Freescale was recently acquired by NXP). However, the marking on the chip is somewhat similar to what Freescale used for HCS08 family of 8-bit microcontrollers. Hence, this area needs to be investigated in more details.

Other components include crystal oscillator with two capacitors, bypassing capacitors, resistors and RF filter with a few inductors and capacitors. The antenna is integrated into the PCB and runs along the edge of the board. The connections to batteries, buzzer and mechanical parts of the Pod are made with springs touching corresponding pads on the PCB.

Also, there is a small detachable corner on the right side through which the wiring to the buzzer is made. This is to allow the users to shut off the buzzer if the device goes into error state with a permanent beeping.

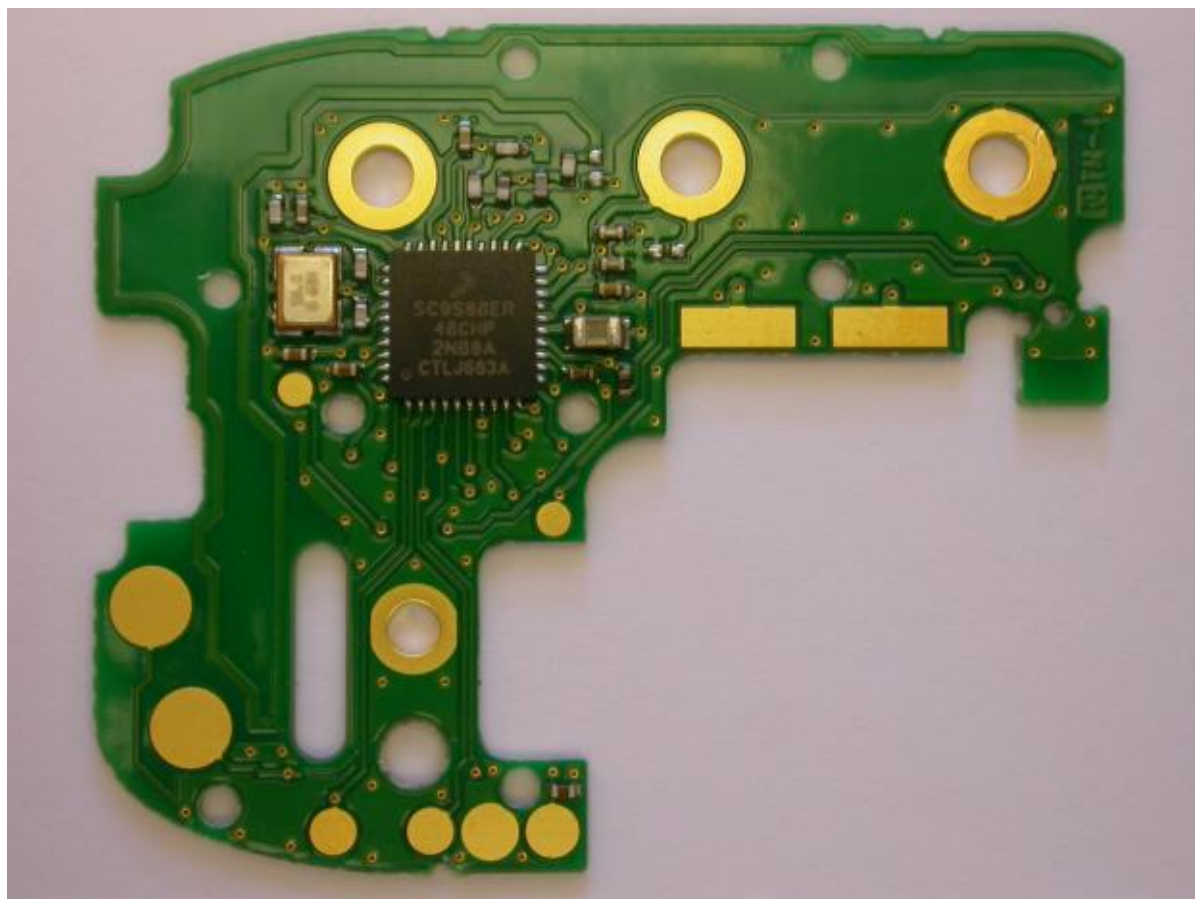

Fig. 10. Components side of the Pod’s PCB.

The connections between the PCB and the mechanical part are shown in Figure 11 with all the spring wires named according to their functionality.

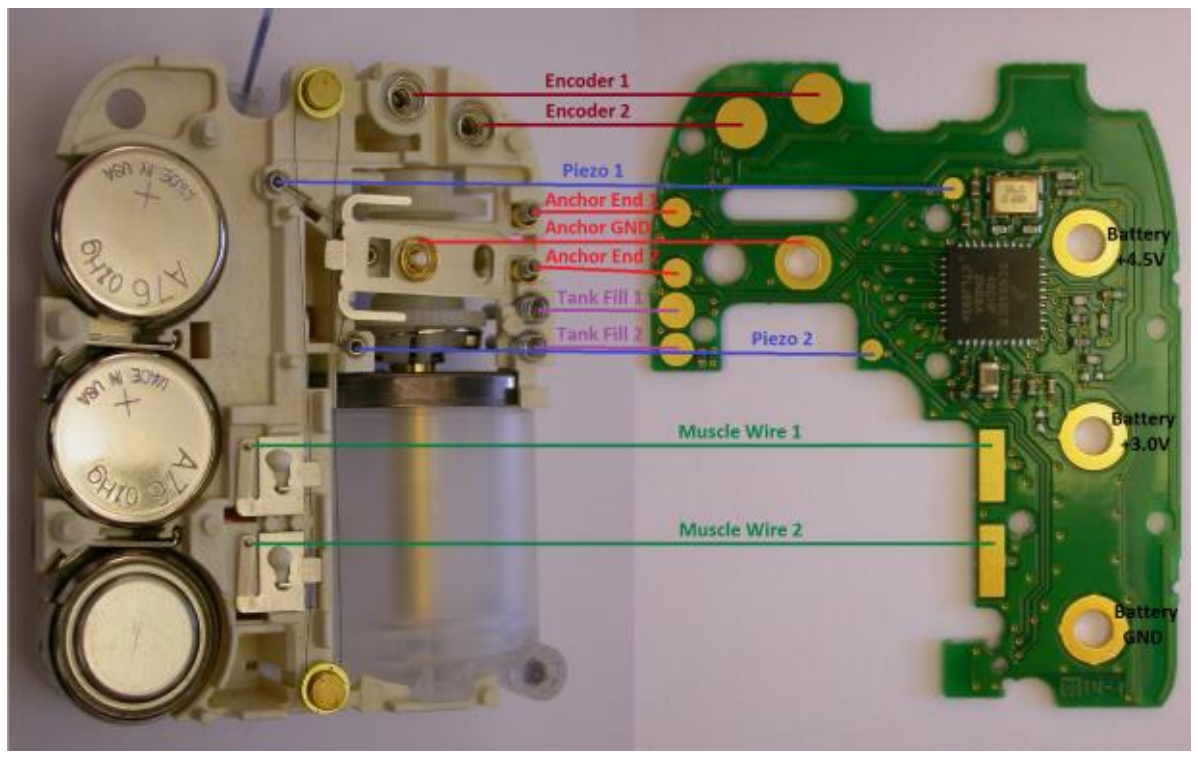

Fig. 11. Connections between the PCB and mechanical part.

## A. *Schematic extraction*

In order to reconstruct the circuit diagram of the device we need to trace all the wires on the PCB that connects all the components together. One possible way of achieving that is to perform an X-Ray imaging. The result of such imaging is presented in Figure 12. This method allows focusing at internal layers of the PCB. That way a full 3D image of the PCB and components could be created.

In case with the Pod there are only 2 layers on its PCB. Therefore, much simpler and less expensive approach can be used. For that all the electronic components were de-soldered using hot air gun at 270°C to heat all parts and then remove them with the help of tweezers. After that the remaining solder was removed with a desoldering braid. Finally the PCB was cleaned with a solvent and dried up. The result of this preparation is presented in Figure 13 with the both sides of the PCB clearly visible and all wires easily trackable.

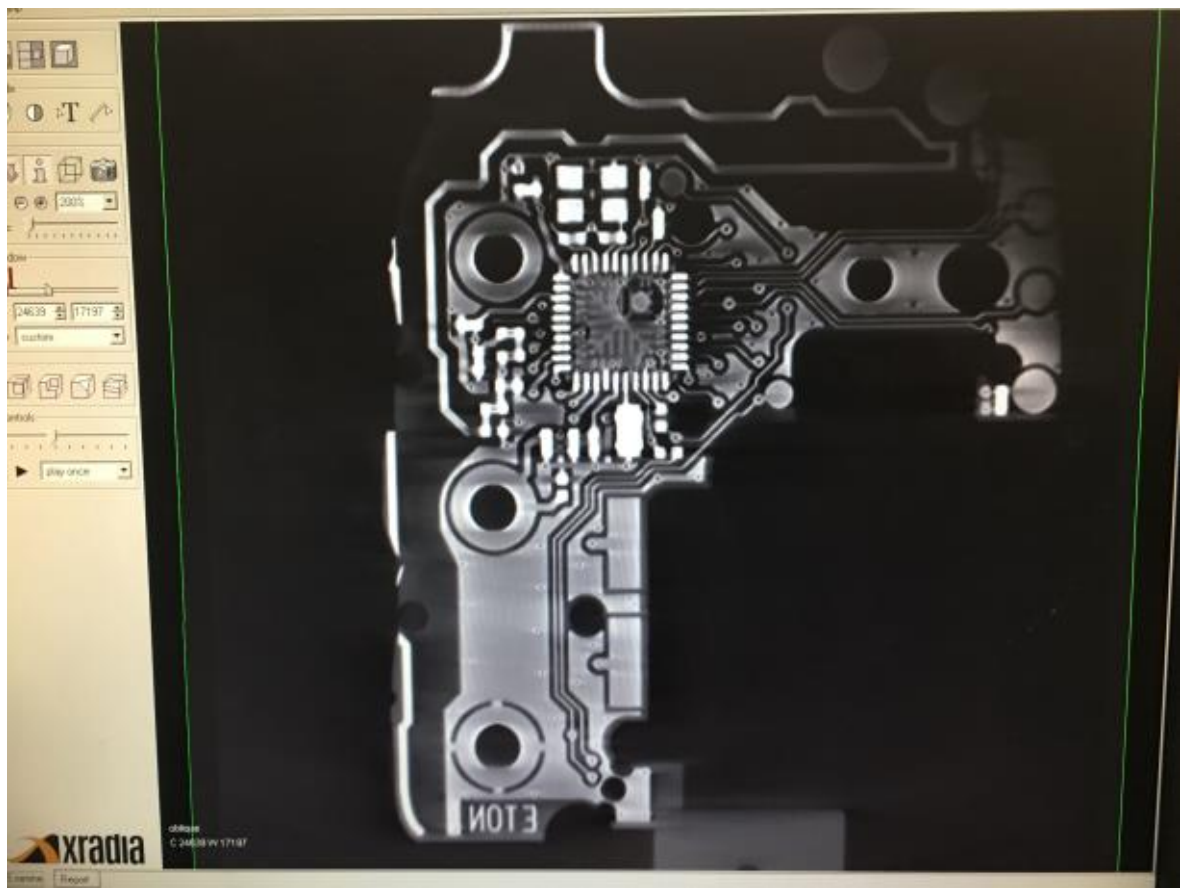

Fig. 12. X-Ray image of the Pod’s PCB.

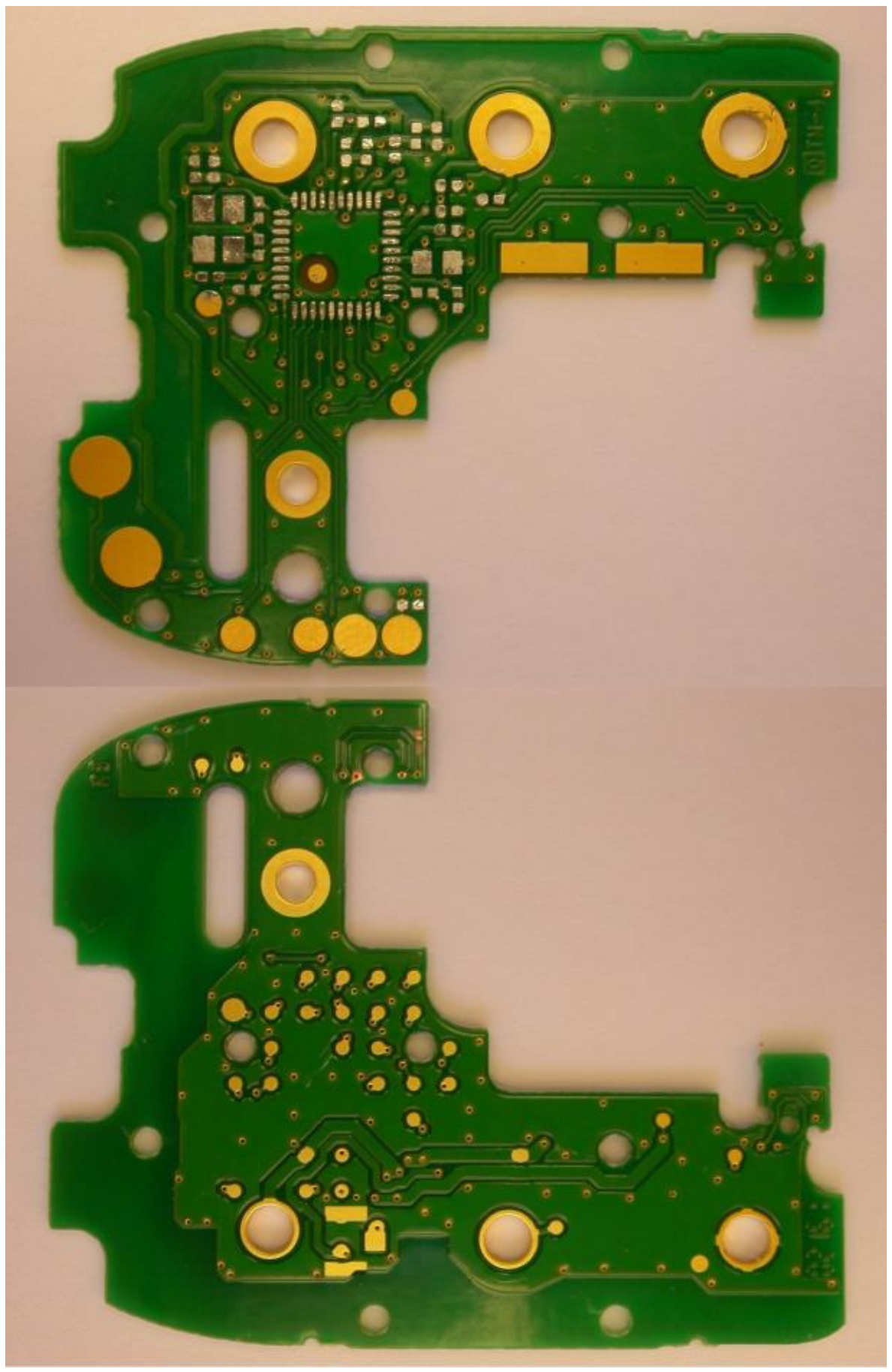

Fig. 13. View on the PCB without components from both sides.

The circuit diagram was manually created from those PCB images, while the actual values of the components such as resistors, capacitors and inductors were measured with an LCR meter. The frequency of the crystal oscillator was marked on it as 26.0MHz. The pinout of the IC was unknown at this stage, hence, some extra work is needed before the complete schematic of the Pod can be created.

### B. *Signal analysis*

The voltages on all SoC pins during the device operation and static parameters of the I/O pins were measured on the core IC of the device. The result is reflected in Table 2.

TABLE II. PINS FUNCTIONS OF THE SOC

| Pin No. | PCB function | PCB level | ΔV GND | ΔV $V_{DD}$ | Type | Comments |
|---|---|---|---|---|---|---|
| 1 | Pad, TP | 0/2.04V | 0.722 | 0.729 | I/O | Encoder |
| 2 | Pad, TP | 0/2.04V | 0.723 | 0.729 | I/O | Anchor T1 |
| 3 | Pad, TP | 0/2.04V | 0.721 | 0.729 | I/O | Anchor T2 |
| 4 | Pad, TP, C | 0/2.04V | 0.726 | 0.726 | I/O | Rod spring |
| 5 | TP | 0V | 0.720 | 0.731 | I/O | Not used |
| 6 | TP | 0V | 0.732 | 0.731 | I/O | Not used |
| 7 | TP | 0V | 0.731 | 0.731 | I/O | Not used |
| 8 | TP | 0V | 0.731 | 0.731 | I/O | Not used |
| 9 | TP | 0V | 0.731 | 0.731 | I/O | Not used |
| 10 | TP | 0V | 0.513 | N/A | I/O | Not used |
| 11 | TP | 0V | 0.732 | 0.730 | I/O | Not used |
| 12 | TP | 0V | 0.726 | 0.724 | I/O | Not used |
| 13 | Pad, TP, R | 0V | 0.732 | 0.728 | I/O | Piezo T1 |
| 14 | Pad, TP, R | 0V | 0.732 | 0.729 | I/O | Piezo T2 |
| 15 | TP, C | 2.87V | 0.352 | $V_{DD}$ | Power | Core Power |
| 16 | C, TP, Battery | 2.87V | 0.449 | 0.132 | Power | Battery 3V |
| 17 | GND | 0V | GND | 0.353 | GND | GND |
| 18 | Pad | 0V | 0.686 | N/A | FET | Motor T1 |
| 19 | Pin 18 | 0V | 0.686 | N/A | FET | Motor T1 |
| 20 | TP, Battery | 4.34V | 0.391 | N/A | Power | Battery 4.5V |
| 21 | Pin 20 | 4.34V | 0.391 | N/A | Power | Battery 4.5V |
| 22 | Pad, TP | 0V | 0.686 | N/A | FET | Motor T2 |
| 23 | Pin 22 | 0V | 0.686 | N/A | FET | Motor T2 |
| 24 | TP | NC | N/A | N/A | NC | Not used |
| 25 | TP, L, C | 2.87V | 0.456 | N/A | RF | Antenna |
| 26 | GND | 0V | 0.690 | 1.044 | RF | Antenna Ref |
| 27 | TP | 0.31V | 0.456 | N/A | RF | Not used |
| 28 | C, TP, Battery | 2.87V | 0.348 | N/A | Power | Battery 3V |
| 29 | GND | 0V | GND | 0.353 | GND | GND |
| 30 | TP, C, C | 0.03V | 0.537 | N/A | Power | Bypass Cap |
| 31 | Pin 30 | 0.03V | 0.525 | N/A | Power | Bypass Cap |
| 32 | GND | 0V | GND | 0.359 | GND | GND |
| 33 | Crystal, C | | 0.669 | 1.098 | Clock | 26MHz |
| 34 | Crystal, C | | 0.655 | N/A | Clock | 26MHz |
| 35 | TP, C | 0V | 0.535 | 1.206 | Power | Bypass Cap |
| 36 | GND | 0V | GND | 0.360 | GND | GND |
| 37 | TP | 0V | 0.718 | 0.728 | I/O | Not used |
| 38 | TP, C | 2.86V | 0.734 | 0.728 | I/O | Not used |
| 39 | TP | 0V | 0.734 | 0.729 | I/O | Not used |
| 40 | TP | 2.86V | 0.734 | 0.729 | I/O | Not used |

TP – test point;
PCB function – connections to the Pin;
PCB level – voltage during operation;
ΔV GND – voltage drop between the Pin and GND line;
ΔV $V_{DD}$ – voltage drop between the Pin and Core power supply;
Type – function of the Pin.

The above measurements allowed narrowing down the number of pins suitable for connecting to the Debug interface which is usually present on HCS08 chips.

During normal operation both Reset and Debug pins must be at logic level '1' (>1.5V). There are only two pins (38 and 40) which satisfy this requirement. These pins were tested with the HCS08 Flash programmer to find the exact locations for the Debug interface. They turned out to be pin 38 for the Reset and pin 40 for the BKGD.

### C. *Wireless communication analysis*

In order to understand how PDM and Pod talk to each other some analysis of their communication is needed. For that a spectrum analyser was used to eavesdrop on the communication during their normal operation. The result of the initial analysis is presented in Figure 14. It can be observed that the communication frequency ranges from 433.8MHz to 434.0MHz.

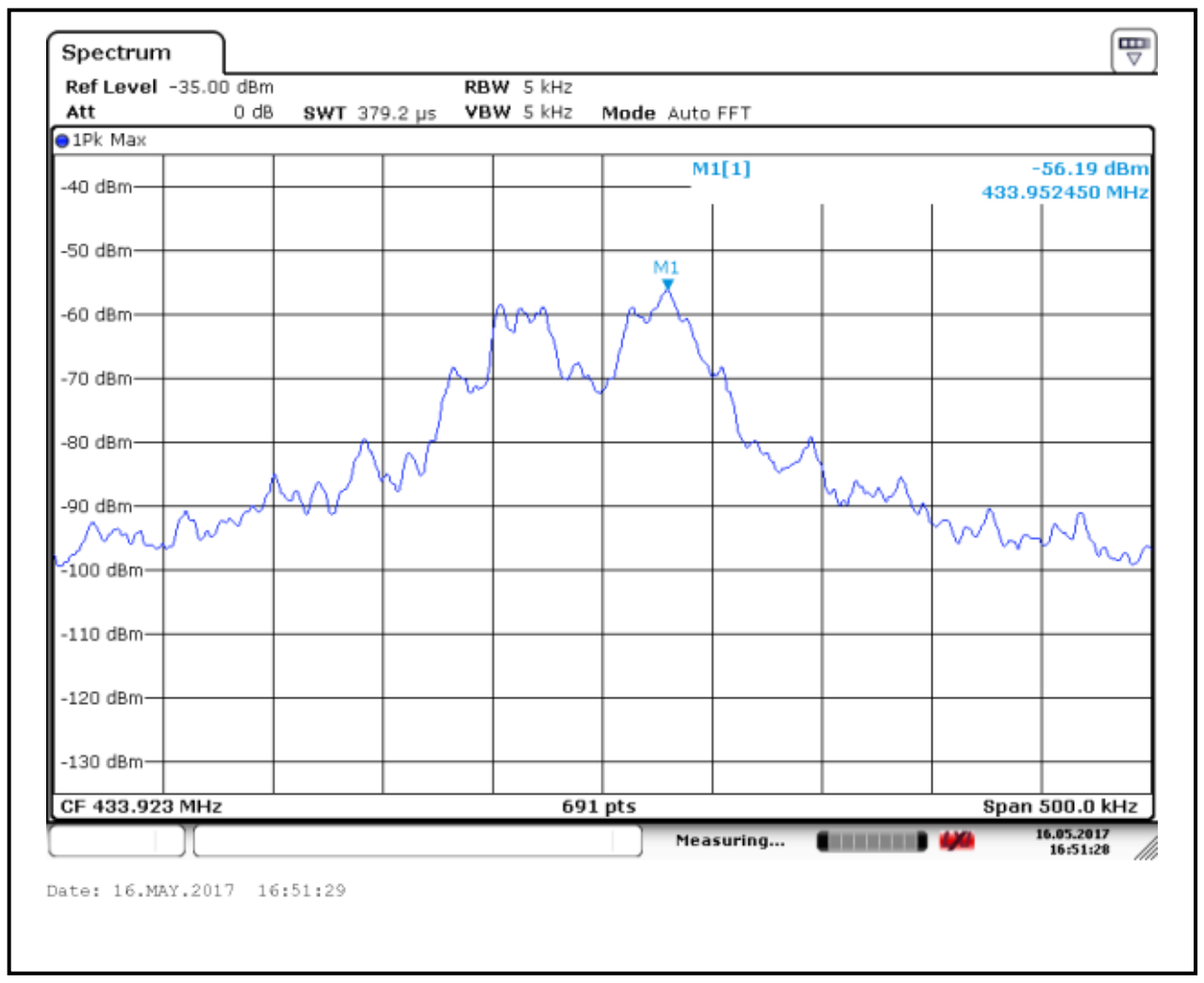

Fig. 14. Spectrum of the wireless communication between PDM and Pod.

The OpenOmni community performed the great job of analysing the wireless communication between PDM and Pod [14]. For the purpose of this teardown there is no need in expanding this area. The communication is based on FSK-2 modulation with 433.9MHz carrier frequency and Manchester encoding. Data bit rate is 40625 baud. Each packet has CRC checksum which was successfully reproduced by the OpenOmni team. However, part of the communication protocol involves some random bytes which they were unable to emulate. Hence, there was the need for more detailed analysis of the Pod's SoC.

## V. COMPONENT IDENTIFICATION AND ANALYSIS

The full name of the core SoC chip is SC9S08ER48CHP.

### A. *Internet search*

Search over the Internet revealed some useful information about other revisions of the chip and some specific parameters. Although there were no direct links to any information on these chips from the NXP website, Google search was able to reveal some hidden pages with specific information about the chip fabrication [17]. It was found that the chip has 48kB of Flash and was fabricated with 0.25µm CMOS process.

### B. *Universal programmers*

Some universal device programmers support this chip. For example, the Elnec BeeProg2 has MC9S08ER48 and SC9S08ER48 in the supported device list [18]. Since the programmer supports QFN package, it can be used to identify the pin numbers for the Reset and Debug with the help of an oscilloscope. However, the ability to program these devices is restricted to only specific serial numbers of the programmer hardware. If anyone wants to use the Elnec

BeeProg2 programmer to Read and Write this chip he has to pay Elnec a few hundred Euros.

The same universal programmer can also be used in ISP mode to program the chip directly on a board. Once the locations of Reset and Debug pins were found it was trivial to connect the programmer and test the chip.

### *C. Decapsulation and deprocessing*

Although the SoC package looks as a QFN it is actually an LGA package with a thin PCB at the base. By careful polishing from the back side the internal wiring of the pins to the bonding pads was revealed. The polishing process exposed four layers on this chip: solder, vias, die and bonding wires (Figure 15). When combined together they reveal the connections between the SoC's silicon die and the LGA-40 package pins.

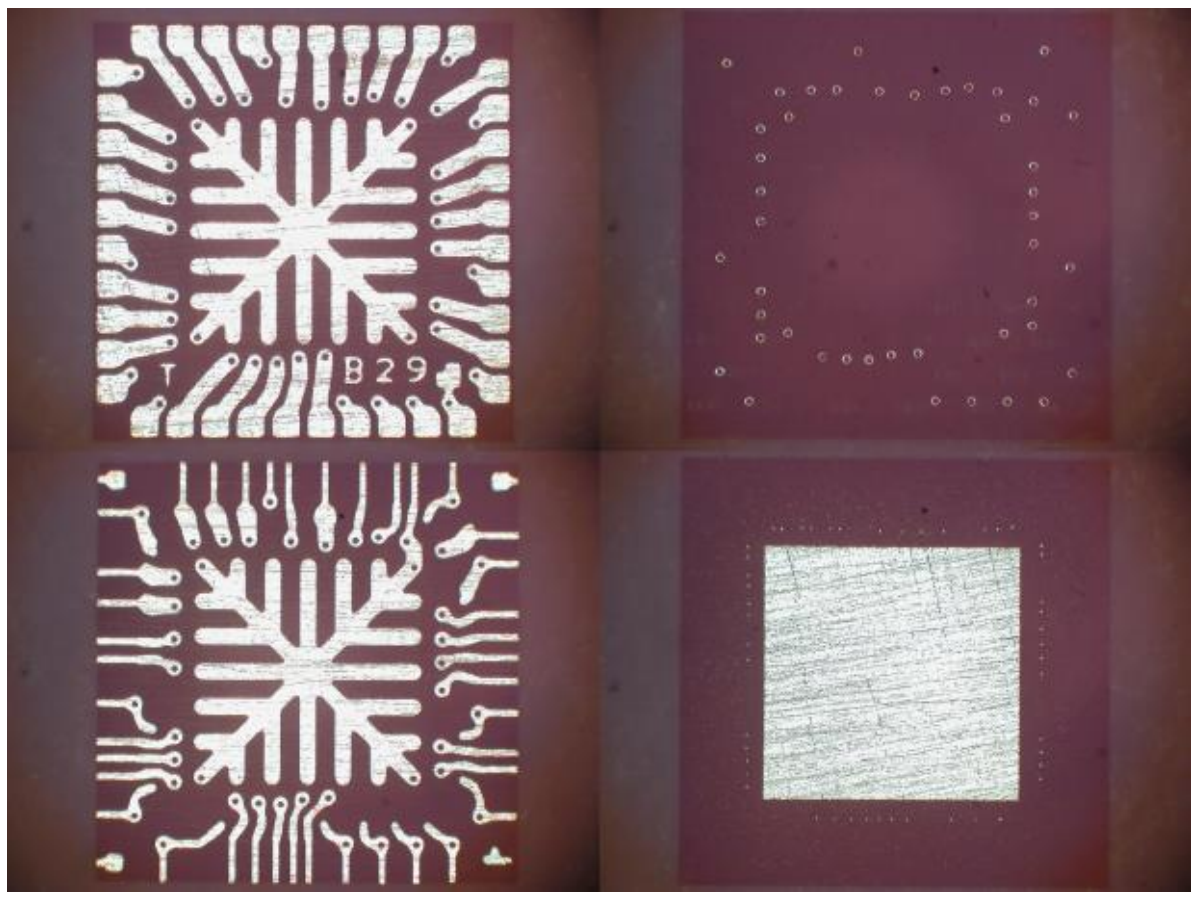

Fig. 15. Results of the SoC's LGA package polishing.

Another way of understanding the internal structure of the chip is to look at its internal surfaces. Fuming nitric acid is usually used for removing the plastic above the chip surface [19]. The result of this operation is presented in Figure 16.

The front side image was used to identify any markings on the surface which could lead to finding more information about the chip. Those markings were: Freescale ©2010, 9S08ER48, N80A, POROSTOP ES2_0 ES2_1, PENIA ES2_0 ES2_1.

In order to observe the internal structure of the chip it needs to be de-processed. This can be done from both its front and back sides. However, the backside de-processing usually gives more information as it allows observation of several layers at a time, while the front side approach would require selective removal of several metal layers. The result of the backside de-processing is presented in Figure 17.

Backside image gave more information about the fabrication process and memory sizes which were confirmed as 0.25μm, five metal layers, 48kB of Flash memory and 4kB of RAM. Half of the chip surface is occupied by the RF communication module.

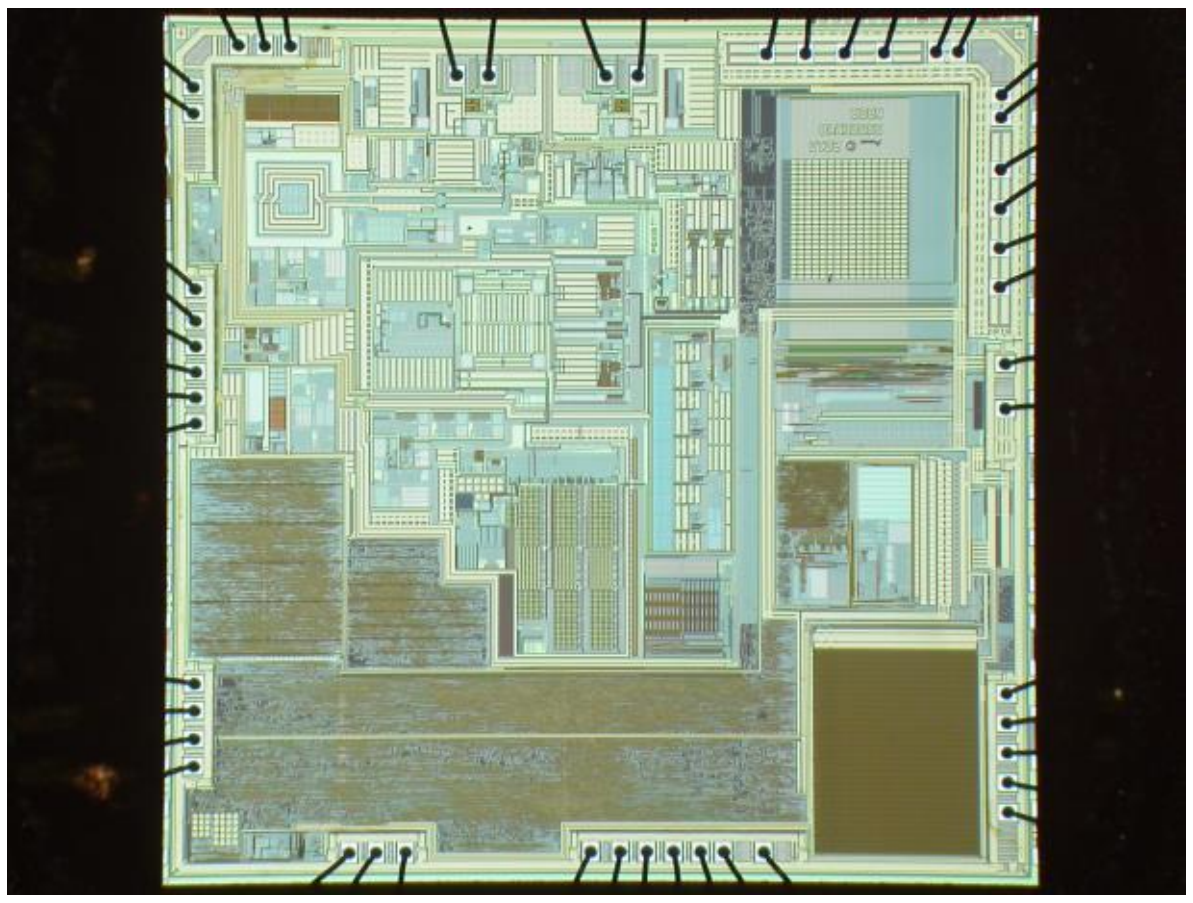

Fig. 16. Front side image of the SoC chip.

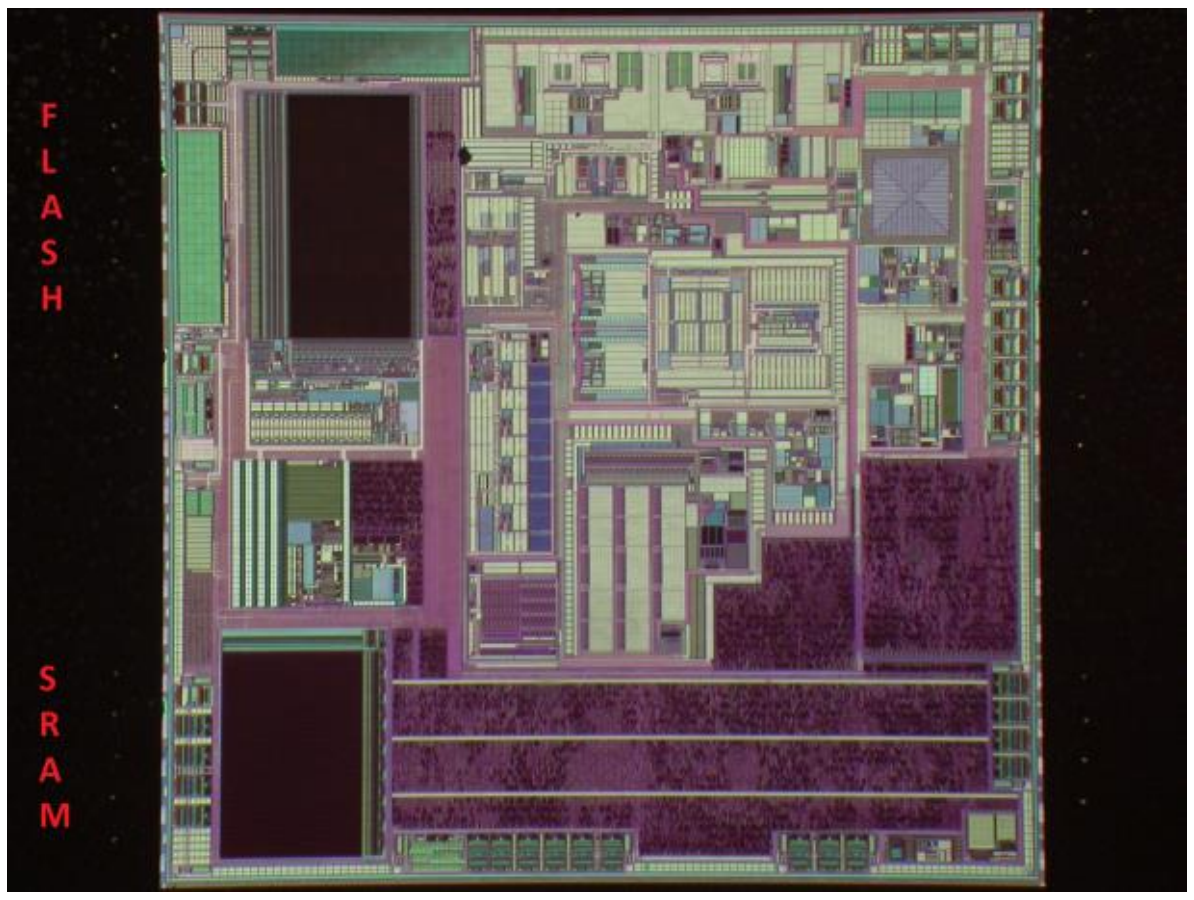

Fig. 17. Backside image of the SoC chip.

### *D. Development tools*

Development tools could contain some crucial information about similar chips in HCS08 family. This information could help in understanding the structure and functions of an undocumented custom chip from the same family.

Freescale provides free development tools for HCS08 family of microcontrollers called CodeWarrior. The latest version as well as some old versions are available on the NXP website [20]. All available versions were downloaded and installed in order to analyse the design support files in the tools directories. Table 3 shows which versions of the CodeWarrior software contain crucial information about the CPU special registers and configuration.

As it can be noticed, only older versions contain all the files, while C and ASM support for 9S08ER48 chip was removed from the version 10.3 and above. Nevertheless, even those newer version contain useful information about special functions and the chip pinout.

TABLE III. FILES PRESENT IN CODEWARRIOR DIRECTORIES

| Ver | Date | **er48.h | **er48.asm | CPU | Configuration |
|---|---|---|---|---|---|
| 5.7.0 | 05/2006 | No | No | No | No |
| 6.3.1 | 03/2010 | Yes | Yes | Yes | Yes |
| 10.1 | 10/2010 | Yes | Yes | Yes | Yes |
| 10.2 | 10/2010 | Yes | Yes | Yes | Yes |
| 10.3 | 08/2012 | No | No | Yes | Yes |
| 10.4 | 04/2013 | No | No | Yes | Yes |
| 10.5 | 09/2013 | No | No | Yes | Yes |
| 10.6 | 02/2014 | No | No | Yes | Yes |
| 10.7 | 02/2014 | No | No | Yes | Yes |

## E. *Circuit diagram*

Now all the necessary information is obtained to complete the schematic of the Pod. For that a free version of Eagle PCB design software was used [21]. The result is presented in Figure 18.

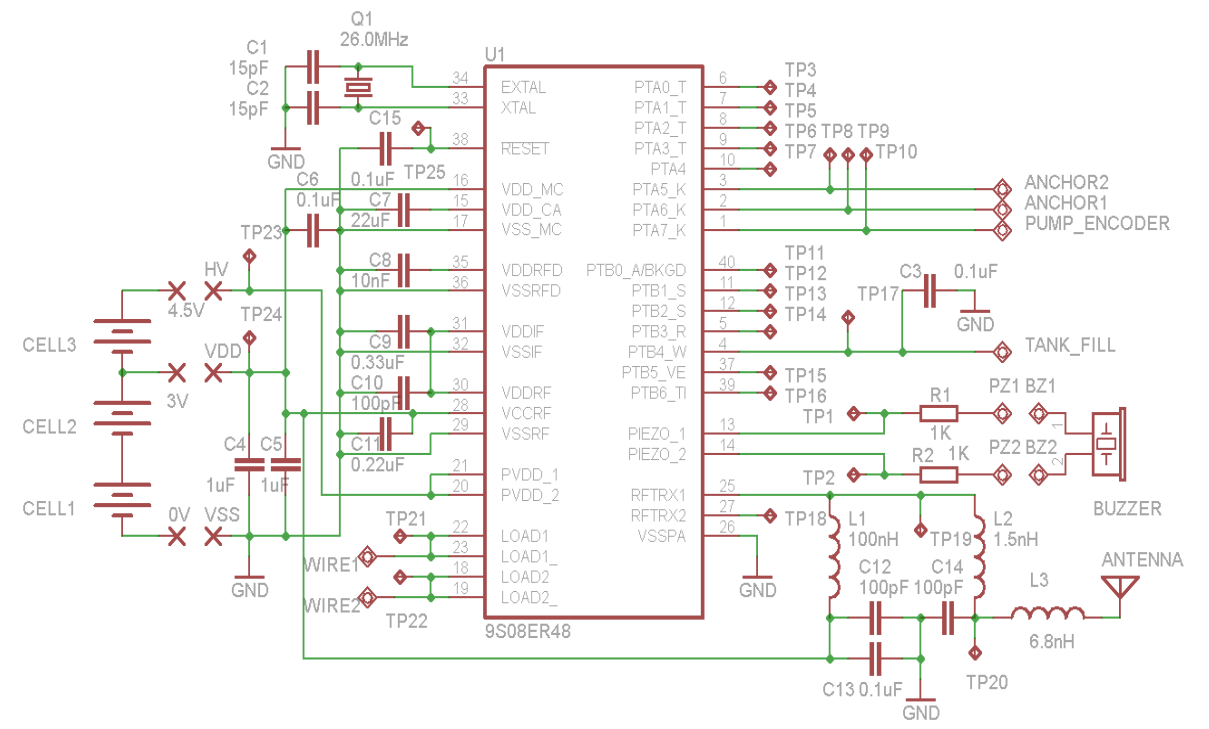

Fig. 18. Circuit diagram of the Pod.

# VI. FIRMWARE EXTRACTION

After some information about the chip was found, in particular the Debug interface pins, we can proceed to the firmware extraction. This could involve many different approaches from simple non-invasive methods, to more sophisticated semi-invasive methods and finally ultimate invasive methods [19].

## A. *Debugging interface and security protection*

The easiest way to access the firmware is to use any existing programming or debugging interface for the on-chip embedded Flash. But usually such access is protected in most production devices. The result of the attempt to access the Pod's SoC using Freescale DEMO9S08QG8 development board [22] is presented in Figure 19. The software popped up a message: "Device is secure. Erase? Yes/No". If the erase operation is performed then the whole Flash will contain only 0xFF value. The security protection in this chip was not possible to defeat with known non-invasive or semi-invasive methods.

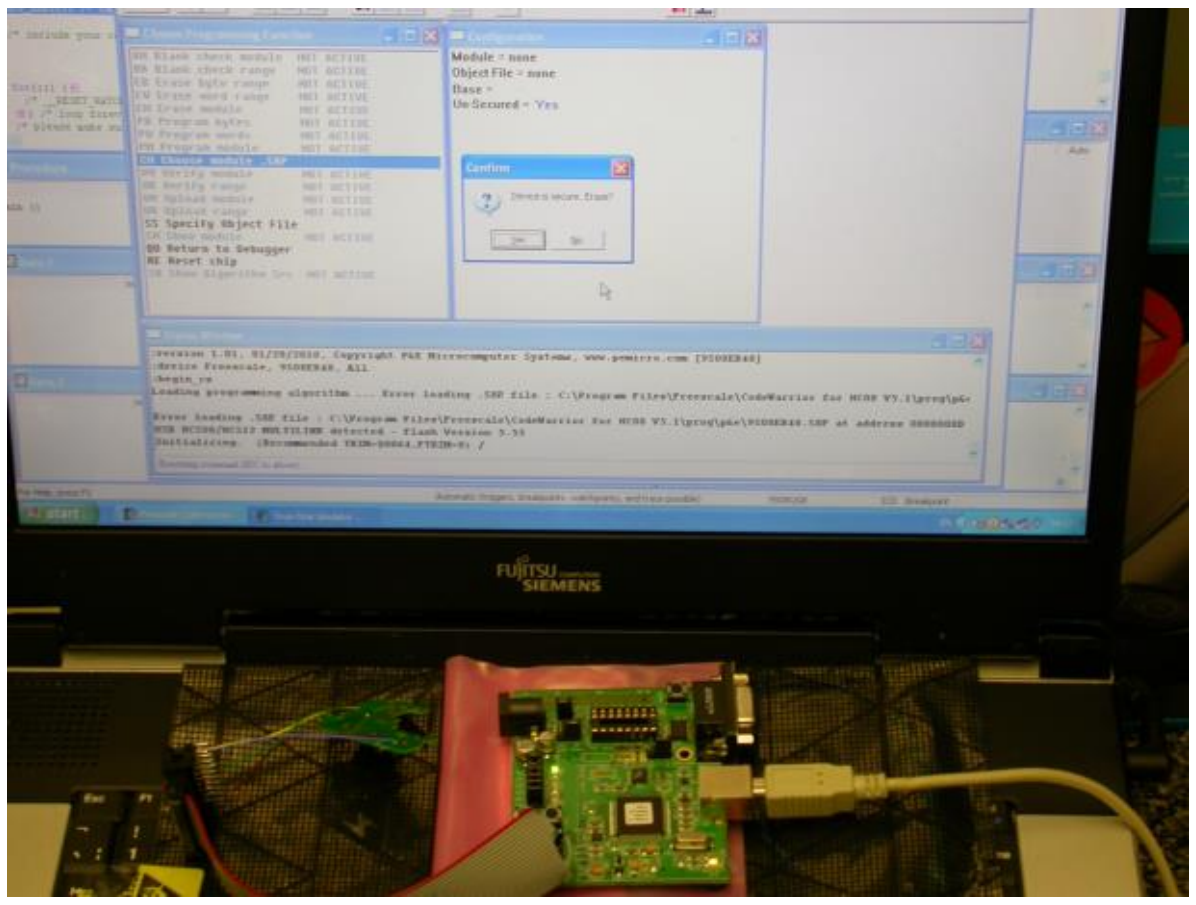

Fig. 19. Accessing the SoC via the Debug interface.

## B. *Invasive attacks*

The SoC chip was fabricated with 0.25µm CMOS process with five metal layers. Wiring the internal data bus to test points on the chip surface would be quite a challenging task. It will not only require Focused Ion Beam (FIB) machine to wire the bus to test points on the chip surface, but also microprobing station to establish connection to them. Although FIB machines are available at many universities for renting at a price starting from $50, their use requires a lot of training and expertise.

## C. *Direct Flash extraction*

There are some known methods of direct data extraction from EEPROM and Flash using Scanning Probe Microscopy [23]. Two mostly used methods are Scanning Capacitance Microscopy (SCM) and Scanning Kelvin Probe Microscopy (SKPM). However, such methods require expensive equipment which is not easy to find for renting. In addition, SCM requires quite sophisticated sample preparation which is extremely difficult to achieve on a single sample for the whole memory array.

Recently introduced direct Flash and EEPROM extraction methods using SEM (Scanning Electron Microscopy) could be used for firmware extraction [24, 25]. However, in order to improve the image quality some additional techniques were used with the help from an industrial collaborator [26]. The example of the resulted image is presented in Figure 20 with brighter areas corresponding to the charged cells which represent 0s.

In order to understand the physical layout of the Flash memory array, some chips were programmed with a test pattern and then imaged to find the corresponding areas on the surface.

In order to image the whole Flash array 96 individual frames were taken and then merged together. Some image processing was applied then to improve the overall contrast. The resulting image was then processed in Matlab. This started with manual pointing at the array corners on the image and manually checking that the grid is correctly positioned over all cells (Figure 21).

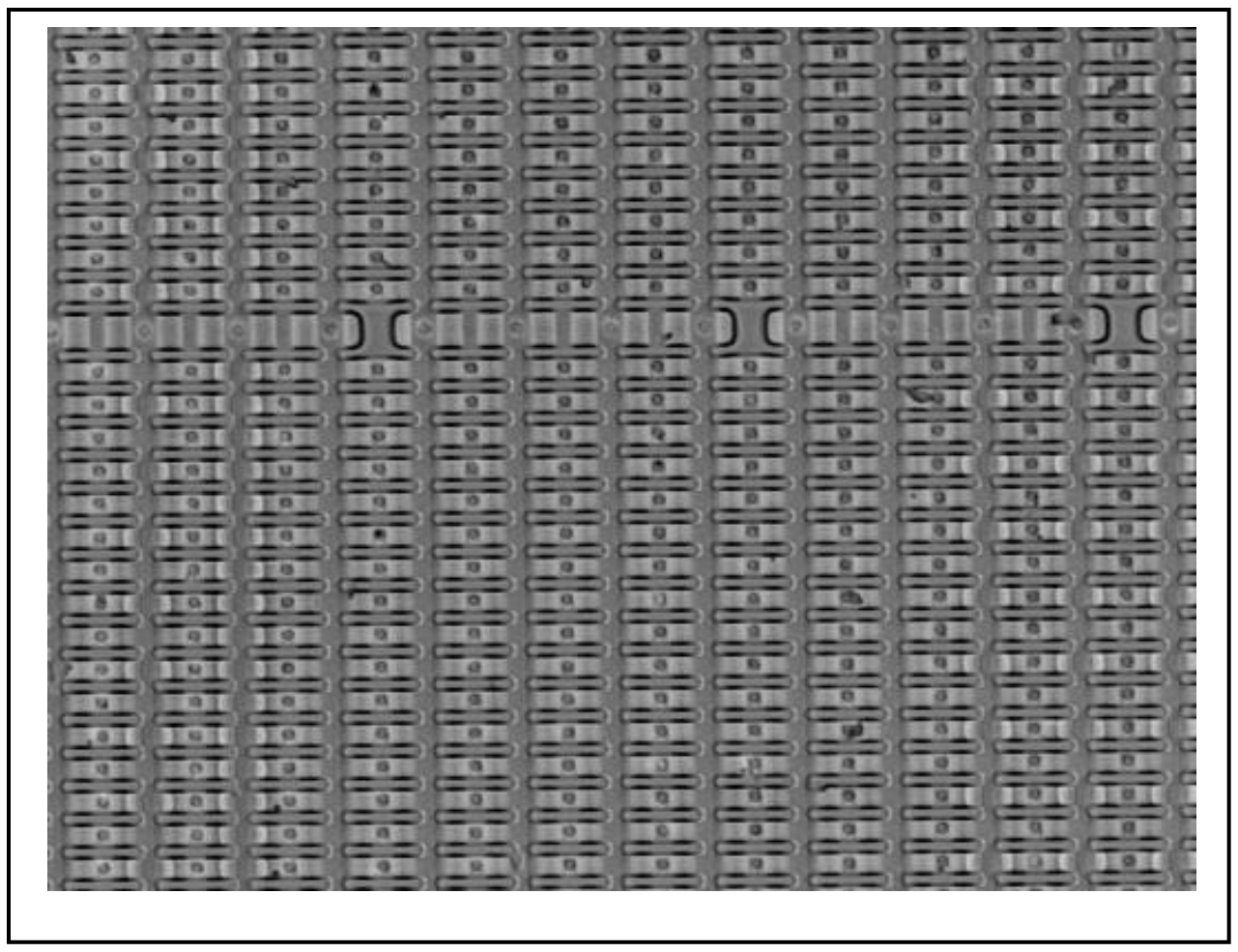

Fig. 20. SEM image of the Flash cells in the SoC.

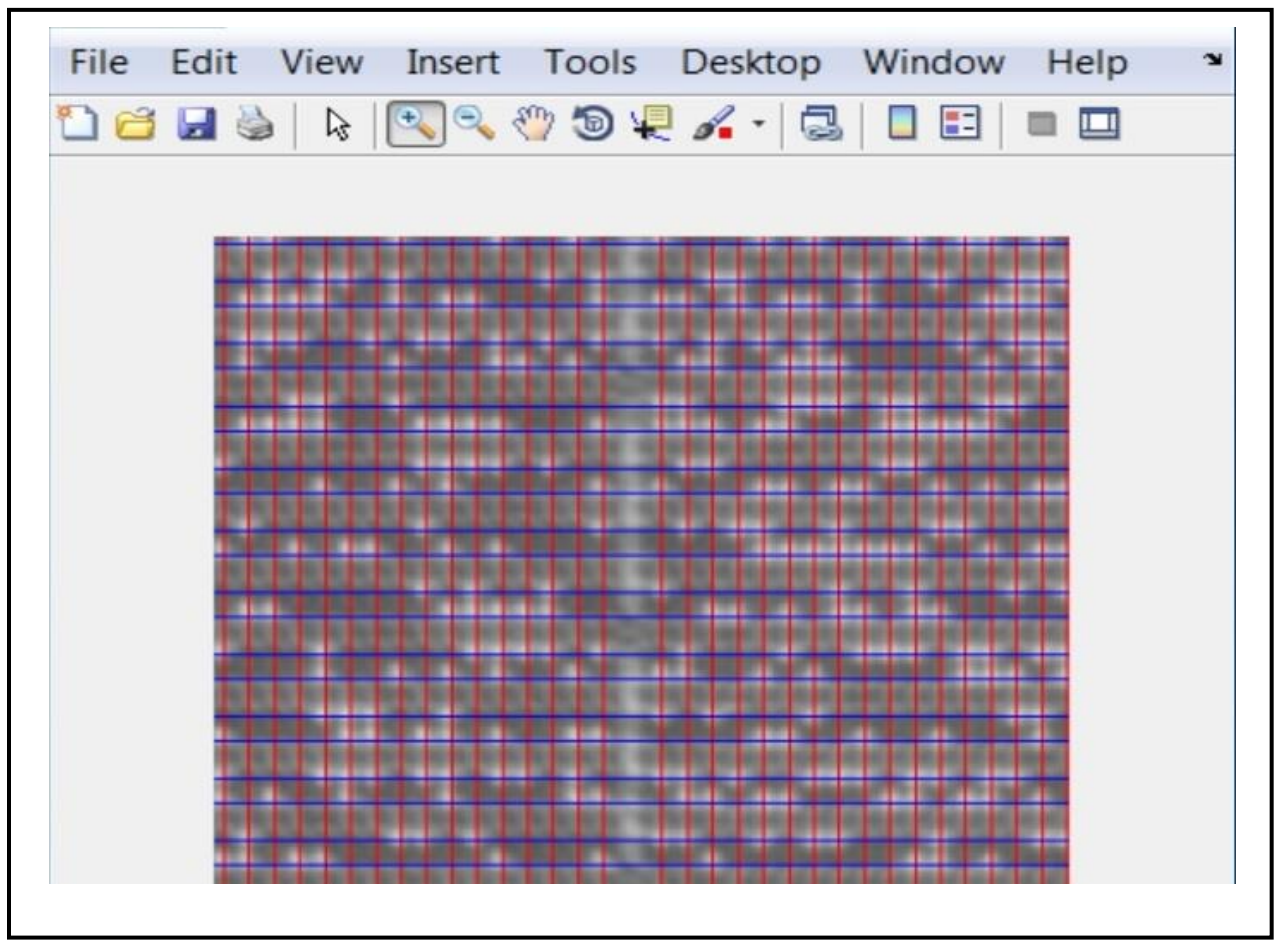

Fig. 21. Checking the grid positioning on the memory array cells.

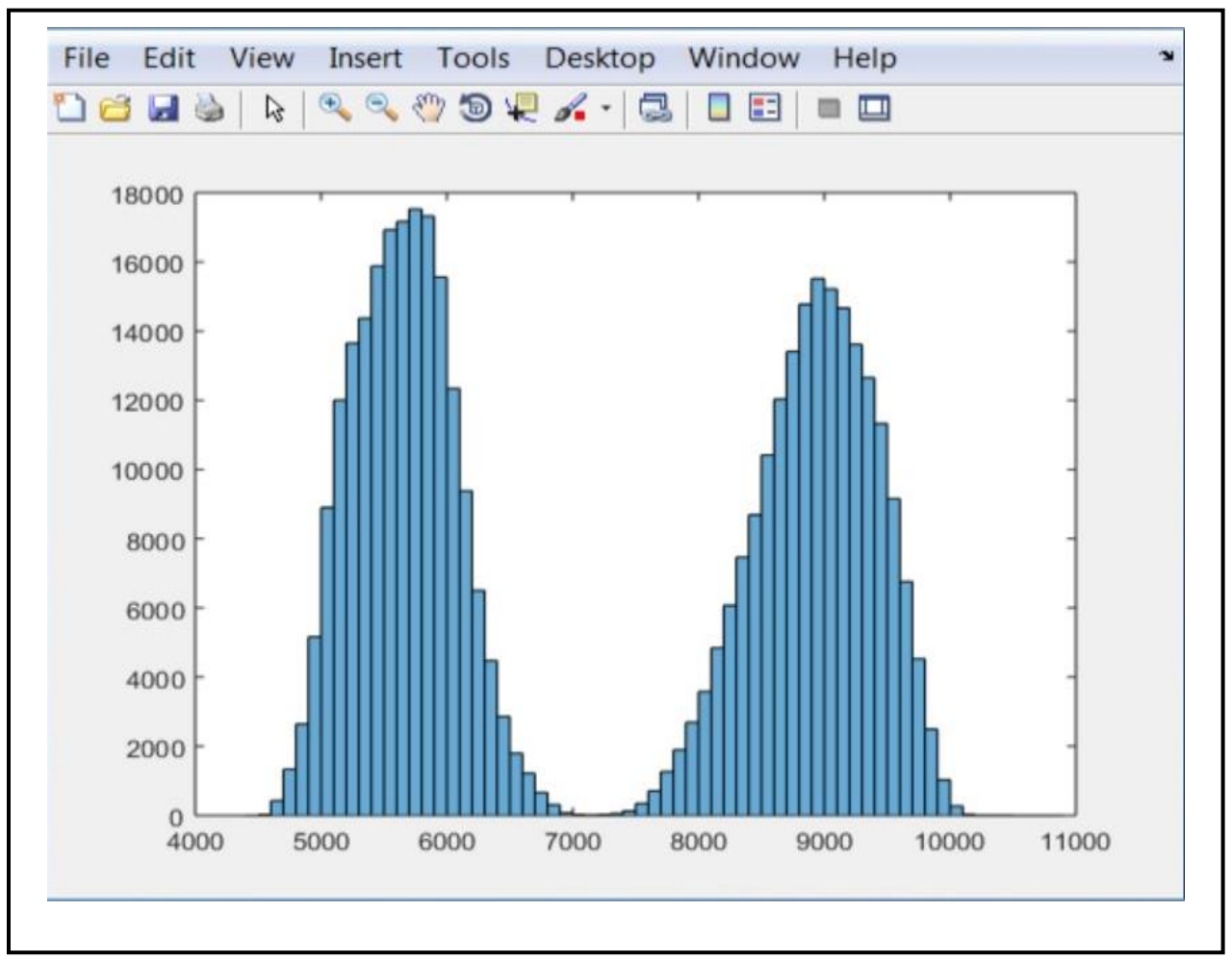

Fig. 22. Histogram of the cells values.

Once the Matlab code has processed the image it shows the histogram of the intensity values acquired over all memory cells (Figure 22). Ideally, there must be a gap between the peaks representing cells with 0s and 1s.

Finally the Matlab code calculates the best threshold value between 0s and 1s and produces HEX file for further firmware analysis and disassembling. If the result is not satisfactory then some adjustments to the image processing or grid positioning can be made. In worst case a new sample chip will have to be prepared and imaged.

To get the best result, several samples of the SoC from new Pods were prepared and imaged. This allowed to produce the final error free HEX file of the firmware for further work. The image processing was also improved with template processing of separate images instead of merging them. This improved the detection rate for 0s and 1s and reduced the error rate to under 0.005%.

### D. *Verification of the firmware extraction*

In order to make sure that the flash contents was extracted correctly, the HEX file was programmed into the used and subsequently deactivated Pod. It worked perfectly and behaved like a brand new Pod. Obviously, the mechanical part cannot be reused because of the cannula insertion mechanism and the lock on the tank membrane rod. However, this proved the full success of the firmware extraction. The video of the whole firmware verification process is available online [27].

## VII. Code Analysis

Code analysis process usually starts with disassembling the code. There are several approaches to address this. The firmware analysis is outside the scope of this paper. Therefore, only recommendations for suitable tools for code analysis are given here.

### A. *Development tools*

Some chips have dedicated debug interfaces, for example, JTAG which allow hardware debugging of the real code inside a physical device. Although the HCS08 core does not have the JTAG interface, it does have a Background Debug mode. The latter can be used to look at the current state of the internal memory and CPU registers. It can also enforce hardware breakpoints either on a specific value of the program counter or on a certain memory or register condition.

### B. *CPU emulators*

Emulators can be used to run the code in real time on a PC or special hardware. These tools are more powerful than debuggers because they are not restricted to limited functions of the device's hardware.

### C. *Professional disassemblers*

One of the best universal professional disassemblers is IDA Pro from HexRays [28]. It supports dozens of different CPU cores and it has an intuitive user interface. It also creates disassembled code which is easy to follow.

The produced HEX file was loaded to verify the code integrity (Figure 23).

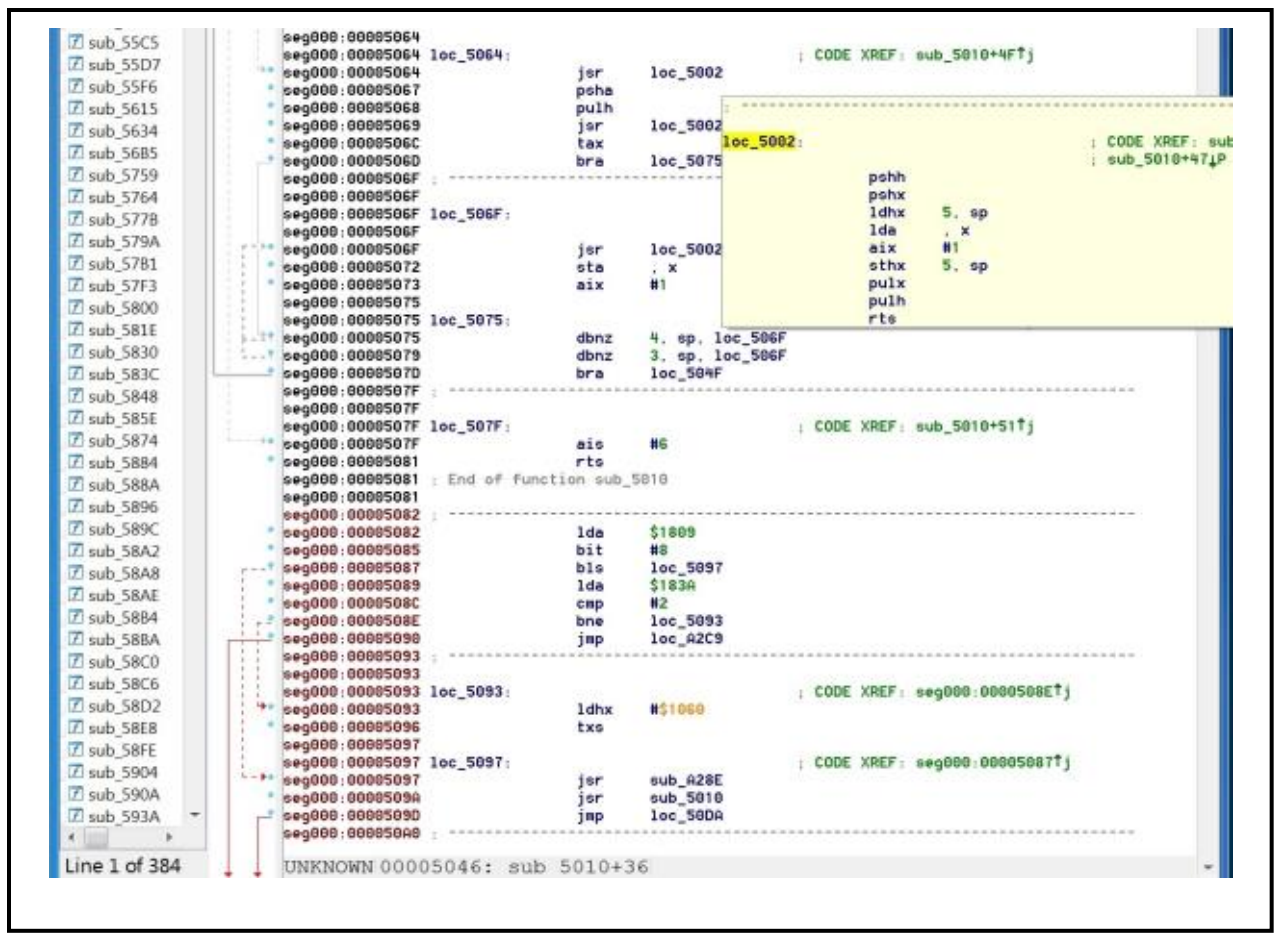

Fig. 23. IDA pro view on the file.

At the first look the code was meaningful which confirmed the successful extraction, but the CPU special registers were not named. In order to improve the readability of the code and help with its understanding, the special configuration file *hcs08.cfg* was created for IDA Pro disassembler. All the necessary information for this file was obtained from *mc9s08er48.h* and *mc9s08er48.inc* files present in one of the directories created by Freescale CodeWarrior development tools. Then the proper assembler code for further analysis was generated from the extracted HEX using IDA Pro disassembler.

### D. Data analysis during the Pod operation

The extracted firmware allowed data analysis inside the Flash memory, because the security protection was not activated during the firmware programming. Hence, it was possible to download the Flash image into a file after running the Pod and then compare it with the original HEX file.

Several areas in the Flash memory were written after the pairing. Table 4 shows the changes to affected addresses. The contents of the Flash memory was also compared with some unique numbers printed on the Pod's case – these are Lot number and TID. They form some kind of a serial number which is unique for each Pod. Quick comparison with the hexadecimal values of the Lot and TID revealed that these values are stored at addresses 4027h–4028h and 4029h–402Ch respectively.

TABLE IV. CHANGES IN THE FLASH MEMORY DURING PAIRING

| Version | Addresses | | | | | | |
|---|---|---|---|---|---|---|---|
| | 4000/1 | 4002/3 | 4008/9 | 4020/1 | 4022/3 | 4024 | 4078 |
| New Pod | FF FF | FF FF | FF FE | FF FF | FF FF | FF | FF |
| Pairing #1 | 1F 03 | 9C ED | FF 00 | 09 07 | 11 0B | 2C | 26 |
| Pairing #2 | 1F 03 | 9C EE | FF 00 | 09 07 | 11 0C | 34 | 28 |
| Pairing #3 | 1F 03 | 9C EF | FF 00 | 09 07 | 11 0D | 06 | 29 |

Address space from 4200h to 4FFFh is used for logging all the events. During the pairing process this area is erased and then written with the proprietary information at addresses 4200h–42FBh. During further communication and insulin delivery process the area is written further with new data.

## VIII. DISCUSSIONS AND FUTURE WORK

The semiconductor device at the core of the Omnipod wireless tubeless insulin pump is a custom made SoC device manufactured by Freescale. There is no directly available information on this device or any form of a datasheet. Also, the standard Freescale/NXP development tools do not support this chip. Nevertheless, some information was obtained from the Internet, development tools and from the analysis of the device. This paved the way to firmware extraction for further code analysis.

One of the possible countermeasures against firmware extraction could be in the use of a memory encryption. However, embedded memory is not particularly suitable for strong encryption. This is because, unlike external memory, it is fetched at random addresses. That poses a big challenge for 8-bit and 16-bit CPU cores [29]. The solution could be in implementing at least 128-bit virtual memory array from which individual bytes are fetched. However, any additional buffers between the memory array and the CPU will increase the latency and reduce the performance.

Future work will involve further analysis of the disassembled code to understand the communication protocol. This work will be carried out by several communities (OpenAPS, Loop, OpenOmni, Nightscout Foundation) to improve the life of patients with Type 1 diabetes. Any further findings, achievements and solutions will be presented by them.

## IX. CONCLUSION

The research presented in this paper shows the overall teardown process carried out on a personal medical device – the Omnipod® wireless tubeless insulin pump. It exposed many challenges addressed during the process and how they were overcome. As a result the complete firmware was extracted without any errors. This paves the way to further analysis of the firmware for compatibility purposes to allow the independent use of the Pods in a closed loop. That way many Type 1 diabetes patients would benefit because at the moment there are no such systems on the market which combine wireless insulin pump and continuous glucose monitor.

All the necessary information about the SoC controller inside the Pod was obtained in the public domain, either through search on the Internet or inside the directories of freely available development tools.

Some security issues were exposed during this teardown process. In order to make personal medical devices more secure manufacturers should improve the hardware security of the semiconductor components which perform all the control, data processing and storage.

This paper exposed some serious security issues associated with the universal development tools. That is the danger of revealing some crucial information about custom and proprietary designs. More care should be taken by chip manufacturers to prevent the leakage of confidential information through their development tools capable of supporting all families of the devices sharing the same CPU core.

## ACKNOWLEDGMENT

I would like to thank Professor Ross Anderson for his help in establishing the legal agreement between the University Research Office and Nightscout Foundation. I would like to thank the Nightscout Foundation for providing donation to cover the cost of components and materials necessary for the sample preparation and analysis. I would like to thank Nanolab Technologies for their help with X-Ray and SEM imaging. I am grateful to CRE Ltd for free academic licensing of their proprietary sample preparation and image enhancement technology. I would like to thank Dr Markus Kuhn for his help in analysing the wireless communication. I would like to thank Joseph Moran, Dan Caron and Pete Schwamb for their helpful and inspiring discussions throughout the project.